\documentclass{article}

\usepackage{iclr2027_conference,times}
\usepackage{amsmath,amssymb}
\usepackage{booktabs}
\usepackage{graphicx}
\usepackage{xcolor}
\usepackage{float}
\usepackage{placeins}
\usepackage[most]{tcolorbox}
\usepackage{CJKutf8}
\usepackage{CJKpunct}
\usepackage[hyphens]{url}
\usepackage{hyperref}
\graphicspath{{figures/}}
\newcommand{\sysname}{\textsc{ReRankID}}

  \newcommand{\codeurl}{\url{https://github.com/lilv98/ReRankID}}
  \newcommand{\codenote}{Code}
  \newcommand{\anonword}{}

\newtcolorbox{promptbox}[1]{breakable, colback=gray!4, colframe=black!60,
  boxrule=0.4pt, arc=1.5pt, left=8pt, right=8pt, top=6pt, bottom=6pt,
  title=#1, fonttitle=\bfseries\small, coltitle=white,
  fontupper=\footnotesize\ttfamily\raggedright}

\title{Reasoning Reranking with Semantic IDs as the Sole Item Representation}

  \iclrfinalcopy
  \hypersetup{pdftitle={Reasoning Reranking with Semantic IDs as the Sole Item
    Representation}, pdfauthor={Zhenwei Tang, Jintao Liu, Di Bai, Xidong Wu,
    Xinwu Cheng}}
  \author{Zhenwei Tang$^{1,2,*}$, Jintao Liu$^{1}$, Di Bai$^{1}$, Xidong Wu$^{1}$,
    Xinwu Cheng$^{1}$ \\
    {\normalfont $^{1}$Google \quad $^{2}$University of Toronto}}

\begin{document}
\maketitle

  \lhead{Preprint}
  {\renewcommand{\thefootnote}{*}\footnotetext{Correspondence to:
    \texttt{josephtang@cs.toronto.edu}, \texttt{lilvjosephtang@google.com}.}}
  \vspace{-2\baselineskip}

\begin{abstract}
Reasoning rerankers write a rationale before ranking a candidate slate,
and existing systems read item metadata to do so, even though generative
recommenders increasingly represent items as compact Semantic IDs (SIDs).
Ranking from SIDs alone has two appeals: SID prompts are far shorter than
the matched item text, and although item codes are fixed from content, the
backbone learns their token embeddings in part from user-item interaction
sequences, so the tokens can carry behavioral
signal that item text lacks.  That training targets next-item generation,
and whether it supports ranking a supplied slate from SIDs alone is
untested.  We build \sysname{}, a
controlled testbed in which models reason and rank from a history and
slate shown only as SIDs, varying item
rendering, history length, and slate size one at a time.  On this
testbed, such models often copy the input order instead
of reranking, and readable metadata does not repair the failure.  We
compare supervised fine-tuning, rejection-sampled self-training,
on-policy distillation, and verifier-based reinforcement learning as
post-training methods and find that they elicit the missing
behavior.  The strongest post-trained model, the 8B model trained with a graded
ranking reward, gives the best SID-only ranking and matches strong text-reading
LLMs whose prompts are about $20\times$ longer.  It ranks above random with
no history at all, where a caption-reading reference does not, and it keeps
its gains at long histories, where text prompts pass $50$k tokens and the
caption-reading reference stops improving.  In a paired rationale audit,
this model argues trade-offs better than that reference but describes
items less accurately.  The common post-training methods we test thus
elicit ranking from SIDs but not faithful item descriptions, which calls
for post-training that improves the reasoning itself, not only the
ranking it produces.
\end{abstract}

\section{Introduction}
\label{sec:intro}

Modern recommenders choose what to show through a cascade of stages, and
the final stage reranks a given slate of candidates for the user.  This
stage depends on detailed comparison against the user's history and among
the candidates, and reasoning rerankers make this comparison
explicit: before the final ranking, they verbalize a rationale a reader
can check
\citep{rankgpt,rankzephyr,first,llmrank_hou}.  In parallel, generative
recommenders increasingly represent items as compact Semantic IDs (SIDs)
\citep{tiger,hua_index,openonerec}.  Two properties of these identifiers
suit reranking.  First, they are short.  A text rendering of a long
history and slate runs to tens of thousands of tokens, which is costly and
is where LLM ranking weakens \citep{lostinthemiddle,llmrank_hou}; the
matched SID rendering stays in the hundreds.  Second, their token
embeddings are learned in part from behavior.  In OpenOneRec, each item's
code is assigned by quantizing a text embedding of the item, and the
backbone then learns the embedding of each code token in a pretraining
stage that aligns it with text and trains it on user behavior sequences,
the stage OpenOneRec credits with encoding collaborative signal
\citep{openonerec}.  The tokens can therefore carry preference signal that
titles and captions do not contain.

\begin{figure*}[t]
  \centering
  \includegraphics[width=\textwidth]{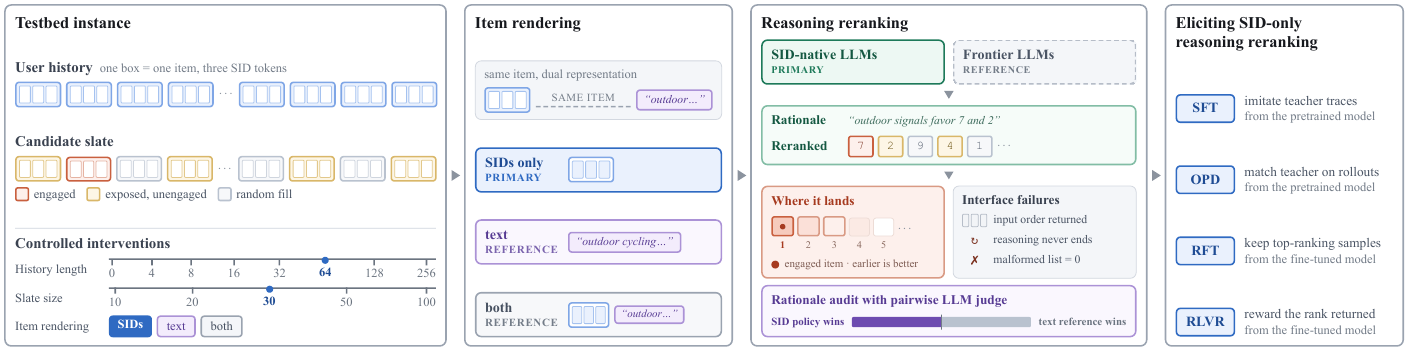}
  \caption{Overview of \sysname{}.  Models read an SID-only history and
  candidate slate, produce a rationale, and return a ranked index list;
  a paired audit compares the emitted rationales.
  SFT and OPD start from the pretrained model, RFT and RLVR continue from
  the SFT checkpoint, and each is compared with its own initializer.}
  \label{fig:overview}
\end{figure*}

Recent SID reasoners align SID tokens with text in one backbone and reason
in language, but they generate the next item from a history rather than
rank a given slate \citep{onerecthink,sidreasoner,openonerec}.  Listwise
reasoning rerankers order an externally supplied slate, but they replace or
augment the identifiers with readable metadata \citep{gr2,diffusiongr2}.
Neither line of work shows whether a model
can rank a given slate of candidates using only what its SID tokens
learned in pretraining.

We study the case between these two lines,
\emph{reason-then-rerank from SIDs alone}.  At inference time, both the user
history and every candidate are rendered only as SIDs; natural-language task
instructions are allowed, but no item text or derived
user profile is shown.  \sysname{} is the testbed we build for this
controlled study (Figure~\ref{fig:overview}).  On the testbed, a model
produces a rationale followed by a ranking of candidate indices.  The
ranking refers to candidates only by index, so the same users and slates
can also be rendered as text alone or as both SIDs and text without
changing the task.  The
user-disjoint testbed changes the item rendering, the history length, and
the slate size along one axis at a time, and scores ranking quality alongside interface
failures such as reasoning loops and copying the input order.  We treat
the rationale as what a reasoning reranker shows its reader and audit its
content directly, without assuming that it causes the ranking.

We first diagnose the existing models before any post-training.  The
main failure is in the interface.  At the anchor setting (a 64-item
history and a 30-candidate slate), both models trail the text-view
references and frequently return the candidate order unchanged, and
readable captions shift their failures without improving the ranking.
The SID prompt is roughly $20\times$ shorter than its text
counterpart and nearly $28\times$ shorter at the longest
history, yet longer histories bring the existing models no sustained
benefit.

We then ask whether common post-training methods can elicit
reasoning reranking from an SID-pretrained model.  We define matched
1.7B and 8B recipes for teacher-supervised SFT, verifier-filtered RFT
\citep{star}, OPD on student-generated behavior \citep{gkd}, and RLVR with a
ranking verifier \citep{grpo}.  SFT alone removes the identity and loop
failures and yields clear ranking gains at both scales.  At 8B, RLVR then improves on SFT with a paired
gain whose 95\% user-paired bootstrap interval excludes zero, and gives the strongest SID-only model, on par with text-reading
LLMs that see prompts about $20\times$ longer; at 1.7B it matches SFT.
The graded ranking reward also beats a binary top-one variant in the
direct paired contrast at 8B.  This 8B RLVR model ranks above random
with no history at all, where a caption-reading reference does not, and
it keeps its gains at the longest histories, where the text reference
stops improving.  A paired rationale audit then shows that the strongest
SID model argues trade-offs better than the text reference, yet misstates
items it has never seen as text.  The common post-training methods we
test thus elicit the preference signal and history evidence needed to
order a slate from SIDs, but not faithful item descriptions.

Our contributions are threefold:
\begin{itemize}
  \item We define pure-SID reasoning reranking and study it on a released
  testbed\footnote{\codenote: \codeurl.} with paired item renderings,
  one-axis history and slate interventions, and interface-failure
  diagnostics.
  \item We diagnose the existing models before any post-training: they
  neither reliably handle the listwise interface nor benefit from longer
  histories, and readable captions repair neither failure.
  \item We compare four common post-training methods under one protocol
  and find that the best post-trained model uses information a text
  reader does not: it ranks above random without history and keeps its
  gains at long histories where a text reference stops improving.  A
  paired rationale audit shows that this ranking ability does not come
  with reliable item descriptions.
\end{itemize}

\section{Related Work}
\label{sec:related}

\paragraph{Semantic IDs and reasoning over them.}
Generative recommenders index items as compact sequences of discrete
Semantic ID tokens and generate the next item directly in this token space
\citep{tiger,hua_index}.  OpenOneRec releases 1.7B and 8B models built on one
SID tokenizer, together with training data and pipeline, which makes
controlled downstream studies possible \citep{openonerec}.  One line
examines the representation itself: attaching SID tokens to a language
model does not by itself make their semantics usable \citep{sidencoder}, and
the tokens' capacity and decoding structure limit what they express
\citep{sidlimits}.  Another aligns SID tokens with text in one backbone and
adds verbalized or staged reasoning before item generation
\citep{onerecthink,onereason,sidreasoner}.  Because the main decision in
these systems remains which catalog item to generate, reasoning quality
and decoding constraints cannot be separated.  We instead fix one existing
model family, remove the generation step, and test whether its pretrained
SID tokens alone support reasoning over a supplied slate.

\paragraph{Reranking with large language models.}
Listwise rerankers prompt or train LLMs to order retrieved documents
\citep{rankgpt,rankzephyr,first}, and recent systems devote explicit
test-time reasoning to this step \citep{rank1,rankk}, which REARANK
trains with reinforcement learning from a listwise NDCG reward
\citep{rearank}.  In recommendation,
LLMs can rerank candidate items without task-specific training, but their
accuracy is sensitive to
history length, candidate-set size, and presentation order
\citep{llmrank_hou,invarirank}; ALRO trains the ranking directly with a
differentiable listwise loss and a permutation-consistency term against
position bias \citep{alro}.  The GR2 family is closest to our setting:
a model generates a rationale and a ranking for a supplied slate \citep{gr2},
with industrial and diffusion variants training the same behavior at
scale \citep{gr2industrial,diffusiongr2}.  Related designs reason
pointwise before scoring each candidate \citep{r2rank} or rerank
generatively without a verbalized rationale \citep{glorank}.  No system in
this line verbalizes its reasoning from item identifiers alone.  Wherever a
rationale appears, readable item text is in the prompt.  Such systems
cannot show whether identifier tokens alone carry preference signal, or
whether a rationale stays accurate once item text is withheld.  In
\sysname{}, SIDs are the sole item representation and matched text views
make metadata access a controlled variable, so both questions can be
measured.

\paragraph{Post-training for reasoning rerankers.}
Standard recipes for eliciting reasoning include rejection self-training
(RFT) \citep{star}, on-policy distillation (OPD) \citep{gkd}, and
reinforcement learning with verifiable rewards (RLVR) \citep{grpo}.
Recommendation adaptations distill rationales into smaller recommenders
\citep{rdrec,slmrec} and shape ranking behavior with rank-aware
rewards \citep{sapo,r2ec}; Rec-R1 optimizes the LLM with GRPO against the
downstream ranking metric itself, including product reranking, and
compares this with SFT, rejection sampling, and DPO on text-based product
search \citep{recr1}, while the industrial GR2 pipelines mix several
recipes \citep{gr2industrial}.  We run SFT, RFT, OPD, and RLVR with matched
settings on one pure-SID task, where items are never shown as text, so that each result, positive or null, can be
judged against the supervision it receives.

\section{The \sysname{} Testbed}
\label{sec:benchmark}

We want a controlled study of two questions: whether the existing models
can rerank candidates from SID tokens alone, and which post-training
methods elicit that behavior.  Answering them requires a resource that
does not yet exist.  The same users and slates must be rankable from SIDs,
from item text, and from both, so that metadata access is a variable
rather than a confound; every other factor must stay fixed while one
intervention varies; and interface failures must be measured apart from
ranking quality, since a model can break the output format while still
holding useful ranking information.  We build \sysname{} to meet these
requirements.

\paragraph{Problem formulation.}
\label{sec:task}
For a user, let $H_k=(h_1,\ldots,h_k)$ be the $k$ most recent history
items and $C=(c_0,\ldots,c_{N-1})$ a slate of $N$ candidates, exactly one
of which is the observed positive $c^{+}$.  An item rendering $\rho$ maps
an item to tokens, and the model reads the rendered pair
$X_{\rho}=\big(\rho(H_k),\,\rho(C)\big)$, with
$\rho\in\{\rho_{\mathrm{sid}},\,\rho_{\mathrm{text}},\,\rho_{\mathrm{both}}\}$.
Given $X_{\rho}$, the model reasons about the user's interests in natural
language and returns a ranking $\pi=(\pi_1,\ldots,\pi_{10})$ of ten
distinct candidate indices $\pi_j\in\{0,\ldots,N-1\}$, most relevant
first.  Ranking quality is scored by the rank of $c^{+}$ in $\pi$.
Because $\pi$ refers to candidates only by
index, the same user, slate, and label can be scored under any rendering.

The three renderings are item-matched views of the same example: SID-only
($\rho_{\mathrm{sid}}$), dense-caption text ($\rho_{\mathrm{text}}$), and
combined ($\rho_{\mathrm{both}}$), which shows each item with both its SID
and its caption.  The primary setting is \emph{pure-SID}: inference uses
$\rho_{\mathrm{sid}}$ alone, so item titles, captions, categories, and
derived user profiles are absent, while the task instruction and the
generated rationale remain natural language.  A gap between two views on
the same users is then an item-rendering effect, not a model-family or
population effect.

\paragraph{Controlled construction.}
\label{sec:construction}
\label{sec:axes}
\label{sec:splits}
\label{sec:coverage}
We instantiate the testbed on Kuaishou short-video interactions and the
OpenOneRec SID vocabulary \citep{openonerec}.  Its SID tokens
(\emph{itemic tokens} in its terminology) quantize text embeddings of item
descriptions, and their embeddings are aligned with text and then
co-pretrained on user behavior sequences.  Its matched public checkpoints,
OpenOneRec-1.7B-pro and OpenOneRec-8B-pro, allow a scale comparison under
one identifier system.
The test set contains $400$ exact-one-positive users, balanced between
explicit-engagement positives (a like, follow, or forward) and long-view
positives, so an end-to-end score never
averages over examples with ambiguous credit.  Each user contributes one
$100$-item priority pool: the positive, clean exposure-hard negatives, and
uniform-random catalog fill, with competing positives excluded rather than
treated as negatives, so an item the user engaged with never enters a
slate as a negative.  Slates take the first $N$ pool items under a fixed
deterministic shuffle, so they are nested and growing $N$ isolates the
effect of added candidates.  Histories use the same nested recent
subsequence for every $k$, so the history curve measures added evidence
for fixed users rather than a population shift.

From the anchor setting, $k{=}64$ and $N{=}30$, we vary one axis at a
time: history length $k\in\{0,\ldots,256\}$, slate size
$N\in\{10,\ldots,100\}$, and the item rendering
(Figure~\ref{fig:overview}).  Two further display orders are reserved for
a display-order audit.  Evaluation users are disjoint from every
post-training pool, and each history temporally precedes its target.  This
blocks instance-level leakage from our own training but keeps
representation-level exposure on purpose, since learning collaborative
structure in the SID token space is the capability under study.
Appendix~\ref{app:benchmark} gives the eligibility rules, label construction,
negative recipe, and split manifests, and discusses pretraining exposure
(Appendix~\ref{app:pretraining-exposure}).

\paragraph{Scoring and diagnostics.}
\label{sec:metrics}
\label{sec:stats}
The primary metric is end-to-end NDCG@5, with invalid outputs scored zero
rather than dropped from a valid-only average, so interface failures count
against a model; NDCG@10, Recall@5/10, and MRR are also reported.  Because each
user has exactly one observed positive, every reported metric is a
function of that positive's returned rank.  This is the standard offline
protocol for implicit feedback \citep{sasrec,bert4rec}: the score asks
whether a model moves the observed item up, not whether the order among
the remaining candidates is right, because no interaction log records
that order.  The audit in Section~\ref{sec:results}
separately judges how rationales argue the comparisons among
candidates.  Alongside
accuracy we log each failure mode the task makes possible.  Identity\%
counts rankings that return the candidate order exactly as shown, the
literal ranking $[0,1,\ldots,9]$: under shuffled candidates this output
passes the format check but makes no ranking decision
\citep{llmsrecsyn,gr2,stopoverthinking}.  Loop\% counts responses whose
reasoning never terminates and instead runs to the token cap.  We also
record final validity and use of the fixed extraction fallback.
Main comparisons are paired over the same $400$ users, summarized by a
user-paired bootstrap interval for each method against its own
initializer; we call a difference significant when its 95\% interval excludes
zero.  Every system uses one fixed prompt and one display order rather
than per-model tuning \citep{rankk,drdt}, and we audit both.  Rescoring under the reserved display orders and under natural rewordings
of the prompt shifts the anchor score by no more than rerunning an unchanged
condition does, and the text reference's top pick is statistically
indistinguishable from uniform over the $30$ display positions.  Exact prompts, parsers, and budgets are in
Appendix~\ref{app:evaluation}, and both audits in
Appendix~\ref{app:permutation}.

\section{Diagnosing the Existing Models}
\label{sec:zeroshot}

We first evaluate the existing models before any post-training.

\begin{table*}[t]
  \centering
  \small
  \setlength{\tabcolsep}{3.0pt}
  \resizebox{\textwidth}{!}{%
  \begin{tabular}{@{}lccccccccc@{}}
    \toprule
    & \textbf{Input} &
    \textbf{NDCG@5} & \textbf{NDCG@10} &
    \textbf{Recall@5} & \textbf{Recall@10} &
    \textbf{MRR} & \textbf{Identity\%} &
    \textbf{Loop\%} & \textbf{Input tokens} \\
    \midrule
    Random                     & --   & .105 & .157 & .190 & .350 & .099 & --   & --   & -- \\
    Caption                    & Text & .184 & .248 & .303 & .503 & .171 & --   & --   & -- \\
    Claude-Sonnet-5            & Text & .210 & .282 & .315 & .540 & .205 & 0.0 & 0.0 & $16{,}722$ \\
    Gemini-3.6-Flash           & Text & .247 & .310 & .378 & .570 & .231 & 0.0 & 0.0 & $16{,}722$ \\
    GPT-5.6-Terra              & Text & .256 & .322 & .398 & .603 & .237 & 0.0 & 0.0 & $16{,}722$ \\
    \midrule
    OpenOneRec-1.7B-pro        & SID  & .114 & .167 & .195 & .365 & .109 & 42.3 & 16.5 & $845$ \\
    OpenOneRec-1.7B-pro        & Text & .115 & .165 & .193 & .350 & .110 & 20.0 & 44.0 & $16{,}722$ \\
    OpenOneRec-1.7B-pro        & Both & .139 & .184 & .230 & .370 & .127 & 30.8 & 41.0 & $17{,}255$ \\
    \midrule
    OpenOneRec-8B-pro          & SID  & .168 & .228 & .268 & .455 & .160 & 20.5 & 37.3 & $845$ \\
    OpenOneRec-8B-pro          & Text & .159 & .226 & .265 & .480 & .151 &  2.3 & 12.3 & $16{,}722$ \\
    OpenOneRec-8B-pro          & Both & .166 & .222 & .255 & .428 & .159 & 13.0 &  6.0 & $17{,}255$ \\
    \bottomrule
  \end{tabular}}
  \caption{Anchor results before any post-training ($n{=}400$, $k{=}64$,
  $N{=}30$).
  ``Both'' shows each item with both its SID and its
  caption.  Input tokens are measured with the OpenOneRec tokenizer, one
  yardstick for every row; the APIs' own tokenizers count these prompts
  higher still (Appendix~\ref{app:diagnostic-results}).  Identity\% and
  Loop\% are the interface diagnostics defined in
  Section~\ref{sec:metrics}.  Random and Caption do not implement the
  reason-then-rank interface, so interface fields do not apply.}
  \label{tab:main-anchor}
\end{table*}

\paragraph{Setup.}
We evaluate the 1.7B and 8B OpenOneRec models on the test set,
reporting the anchor setting here and the interventions in
Section~\ref{sec:scaling-results}.  The two models
are scored under all three renderings: SID-only is the primary condition,
and the text and combined rows rerun the same users with readable item
text.  SID inference first permits free reasoning; a fixed
forced-extraction stage is invoked only when no valid ranking is
produced, and a response still invalid after that stage scores zero.
Full protocol constants are in Appendix~\ref{app:metrics}.

\begin{figure*}[t]
  \centering
  \includegraphics[width=\textwidth]{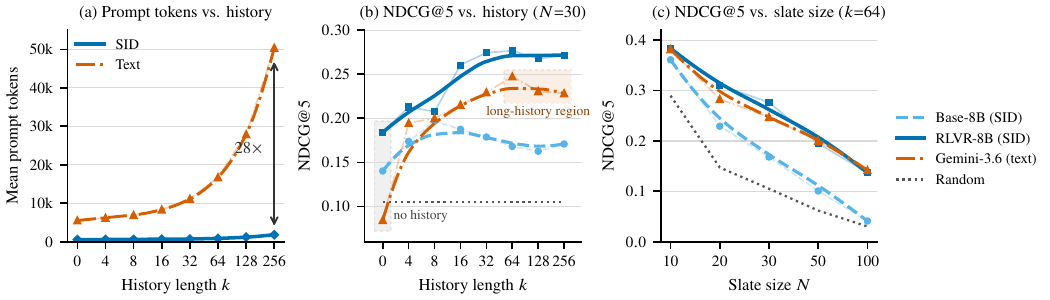}
  \caption{History and slate interventions.  (a)~Mean prompt tokens
  for the SID and text renderings as history grows, measured with the
  OpenOneRec tokenizer on the same users and slates.
  (b)~NDCG@5 as history lengthens ($N{=}30$) and (c)~as candidates are added
  ($k{=}64$), each varying one axis from the anchor setting, for the 8B
  Base, its RLVR model, the Gemini-3.6 text reference, and seeded
  Random.  Shaded boxes mark the no-history setting and the long-history
  region discussed in the text.  Markers are measured values; curves are
  smoothed trends.}
  \label{fig:scaling}
\end{figure*}

\paragraph{Reference rows.}
The other rows of Table~\ref{tab:main-anchor} are reference points.
Random is a fixed seeded shuffle and sets the chance level.  Caption ranks
candidates by embedding similarity to the centroid of the history
captions, a simple non-generative text baseline.  GPT-5.6-Terra,
Gemini-3.6-Flash, and Claude-Sonnet-5
\citep{gpt56,gemini36flash,claudesonnet5} read the item-matched text view
under the same task wording and answer format, and show what strong
general models achieve when readable metadata is available.  Their models
and inputs both differ from OpenOneRec, so they serve as diagnostic
references rather than representation-controlled competitors.

\paragraph{Ranking accuracy.}
At the anchor setting, the 8B model beats seeded Random from SID-only
inputs with a significant paired margin ($+.063$, CI $[+.027,+.099]$), while
the 1.7B margin stays within noise ($+.009$, CI $[-.024,+.041]$); both
remain below every text-view reference (Table~\ref{tab:main-anchor}).
With no history at all ($k{=}0$, Figure~\ref{fig:scaling}b), the 8B
margin over Random from the SID rendering of the candidates alone is
$+.035$ (CI $[-.001,+.071]$), while the Gemini text reference is
statistically indistinguishable from Random; post-training turns this
margin into a significant one (Section~\ref{sec:elicit}).  Longer histories,
in contrast, do not help.  SID prompts stay a small fraction of the
matched text prompts (Table~\ref{tab:main-anchor}, rightmost column), so
extra history is nearly free for an SID model, yet as history grows the
8B model drifts with no sustained trend (Figure~\ref{fig:scaling}b).

\paragraph{Interface failures.}
The main obstacle is the interface, which is new to these models.  No
OpenOneRec training task presents a candidate slate and asks for an
ordering.  The models return the literal input order for a fifth to
nearly half of users, through two failure modes
(Appendix~\ref{app:qualitative}).  In the first, the model closes an
empty reasoning block and emits the input order; in Case~1 the 8B model
stops after $46$ tokens with $[0,1,\ldots,9]$, dropping a positive shown
at position $29$.  This is the usual response of the 1.7B model, for
$71\%$ of anchor users.  In the second, the model transcribes the
prompt's history SIDs in order until the free-reasoning token cap
(Case~2).  This happens for $37\%$ of 8B anchor users, and forced
extraction then mostly returns the input order, giving $75$ of the 8B
model's $82$ identity rankings.  The 8B model does write a nonempty,
terminated rationale for $59\%$ of users, yet ranks only modestly above
Random.  The
two-stage protocol still yields a valid list for every SID anchor
example, but the extraction stage is invoked for $17.5\%$ of anchor users
at 1.7B and $37.5\%$ at 8B, and scoring the free-reasoning stage alone
would lower the SID Base scores from $.114$ to $.101$ and from $.168$ to
$.113$ (Appendix~\ref{app:metrics}).  Readable metadata does not remove
the obstacle.  Given captions instead of SIDs, the same models gain no
consistent ranking accuracy (Table~\ref{tab:main-anchor}): the 8B scores
stay within a few points across all three views, and the 1.7B model
mostly trades identity rankings for reasoning loops.

\section{Eliciting SID-Only Reasoning Reranking}
\label{sec:method}
\label{sec:experiments}
\label{sec:results}

We then ask which post-training methods elicit reasoning reranking, and
how the post-trained models respond to the history and slate
interventions.  We name each post-trained model by its scale and last training stage;
for example, 8B RLVR is OpenOneRec-8B-pro trained with SFT and then RLVR.

\begin{table*}[t]
  \centering
  \small
  \setlength{\tabcolsep}{4.0pt}
  \resizebox{\textwidth}{!}{%
  \begin{tabular}{@{}lllcccccc@{}}
    \toprule
    \textbf{Method} & \textbf{Init} & \textbf{Supervision} &
    \textbf{NDCG@5} &
    $\boldsymbol{\Delta}$\textbf{Base} &
    $\boldsymbol{\Delta}$\textbf{Init} &
    \textbf{95\% CI vs. Init} & \textbf{Identity\%} &
    \textbf{Loop\%} \\
    \midrule
    \multicolumn{9}{@{}l}{\textit{OpenOneRec-1.7B-pro}} \\
    Base & -- & -- & .114 & -- & -- & -- & 42.3 & 16.5 \\
    SFT  & Base & Teacher traces & .225 & +.111 & +.111 & $[+.071,+.150]$ & 0.0 & 0.0 \\
    OPD  & Base & Teacher distributions & .195 & +.082 & +.082 & $[+.044,+.120]$ &
      0.0 & 0.0 \\
    RFT  & SFT & Filtered rollouts & .223 & +.109 & -.002 & $[-.028,+.023]$ &
      0.0 & 0.0 \\
    RLVR & SFT & NDCG@5 reward & .231 & +.117 & +.006 & $[-.015,+.026]$ &
      0.0 & 0.0 \\
    RLVR & SFT & Hit@1 reward & .232 & +.118 & +.007 & $[-.014,+.028]$ &
      0.0 & 0.0 \\
    \midrule
    \multicolumn{9}{@{}l}{\textit{OpenOneRec-8B-pro}} \\
    Base & -- & -- & .168 & -- & -- & -- & 20.5 & 37.3 \\
    SFT  & Base & Teacher traces & .237 & +.068 & +.068 & $[+.028,+.109]$ & 0.0 & 0.0 \\
    OPD  & Base & Teacher distributions & .210 & +.042 & +.042 & $[+.001,+.083]$ &
      0.0 & 0.0 \\
    RFT  & SFT & Filtered rollouts & .211 & +.043 & -.026 & $[-.047,-.005]$ &
      0.0 & 0.0 \\
    RLVR & SFT & NDCG@5 reward & .277 & +.109 & +.040 & $[+.006,+.076]$ &
      0.0 & 0.0 \\
    RLVR & SFT & Hit@1 reward & .232 & +.064 & -.004 & $[-.025,+.016]$ &
      0.0 & 0.0 \\
    \bottomrule
  \end{tabular}}
  \caption{SID-only post-training at the anchor setting.  Each block uses
  its own Base, and each supervision form is defined in
  Section~\ref{sec:controls}.  Confidence intervals are paired bootstrap
  intervals for NDCG@5
  relative to its own initializer.  Repeated RLVR rows compare the
  graded and binary rewards.  Identity\% and Loop\% are the
  interface diagnostics of Table~\ref{tab:main-anchor}; NDCG@10,
  Recall@5/10, and MRR for every model are in Table~\ref{tab:full-anchor}.}
  \label{tab:posttrain-anchor}
\end{table*}

\subsection{Post-Training Setup}
\label{sec:spectrum}
\label{sec:controls}

We compare four post-training methods, each measured against its
own initializer.  SFT and OPD start from Base, while RFT and RLVR continue
from the SFT checkpoint.  They differ in where responses originate and how feedback
enters: SFT imitates a text-view hindsight teacher, whereas OPD, RFT, and
RLVR train on responses sampled by the SID student
(Figure~\ref{fig:overview}, right).  SFT supervises the whole teacher
trace, and the verifiers in RFT and RLVR score the final index list.  We keep the rationale because it is what
reasoning rerankers show a reader, and we evaluate its content directly
(Section~\ref{sec:rationale-quality}).  Training users are
disjoint from the evaluation users.  Within each method, both scales share
the data distribution, prompt, output format, and per-update
optimization settings; all methods use LoRA and leave SID embeddings
unchanged.  The paired intervals in Table~\ref{tab:posttrain-anchor}
resample the evaluation users for each trained checkpoint.  Exact recipes,
training objectives, and compute are in Appendices~\ref{app:posttrain}
and~\ref{app:artifacts}.

\paragraph{SFT.}
\label{sec:sft}
SFT tests whether caption-grounded hindsight demonstrations can teach an
SID-only student usable ranking behavior.  Gemini-3.6-Flash receives the
caption rendering and the known positive, then writes a concise rationale
and a ranked list; the student imitates that response from the
item-matched SID rendering with next-token likelihood \citep{rdrec}, over
$11{,}031$ training demonstrations from users disjoint from evaluation.
Quality gates enforce the output format, positive-at-rank-one
correctness, and candidate grounding, and answer-revealing phrasing is
prohibited and audited.  The teacher rationale is thus a hindsight
training trace, and its ordering of the negatives is teacher-supplied.

\paragraph{RFT.}
\label{sec:rft}
RFT amplifies successful responses that the SFT student already samples.
For each SID-only prompt it draws eight responses, retains
at most the first valid nonidentity trace that places the positive first,
and continues likelihood training from the SFT checkpoint; this is offline
verifier-filtered self-training in the style of STaR \citep{star}.

\paragraph{OPD.}
\label{sec:opd}
OPD transfers reranking from a task-adapted model to the Base model
through dense guidance on states the student itself visits \citep{gkd}.  A fresh Base LoRA samples a response; the frozen SFT teacher
supplies a next-token distribution on each student prefix; the update
minimizes forward KL on the renormalized teacher top-32 support.  Neither
model sees captions, gold positives, or any reward.

\paragraph{RLVR.}
\label{sec:rlvr}
RLVR tests whether the observed positive and a strict output format
provide enough direct learning signal.  Starting from SFT, GRPO
\citep{grpo} samples SID-only responses and assigns
$R(y)=1/\log_2(1+r)$ when $y$ is valid and returns the positive at rank
$r\leq 5$, with ranks counted from one, and zero otherwise; a valid response must close its reasoning
block and emit exactly ten distinct in-range indices.  The reward
equals the primary NDCG@5 metric, with no auxiliary shaping,
and training runs $100$ optimizer steps at each scale.  A controlled
ablation replaces this reward with binary top-one correctness under the
otherwise identical recipe.

\subsection{Results}
\label{sec:elicit}

\paragraph{Main results.}
Table~\ref{tab:posttrain-anchor} separates the two scales and every
initializer.  SFT already teaches the interface: at both scales it
eliminates every identity permutation and reasoning loop and adds a large,
significant ranking gain over Base.  The extraction fallback nearly
disappears with it; across the SFT, RFT, and RLVR models it is invoked for
at most $1.0\%$ of anchor users and adds at most $+.004$ NDCG@5.  The SFT
anchor estimates come close to the Gemini text reference, and neither the
remaining gap to Gemini nor the gap between the two SFT scales is
significant.

RLVR is the only method that improves on SFT, and only at 8B, where the
graded ranking reward adds a significant paired gain and yields the strongest SID-only
anchor estimate, $.277$, on par with the Gemini and GPT-5.6-Terra text
references that read prompts about $20\times$ longer (paired differences
$+.029$ and $+.021$, Appendix~\ref{app:anchor-results}).  In hit-rate
terms, it places the positive among its returned ten for $64.0\%$ of
users, against $57.0\%$ for Gemini and $60.3\%$ for GPT-5.6-Terra
(Tables~\ref{tab:main-anchor} and~\ref{tab:full-anchor}).  At 1.7B the same
recipe stays within noise of SFT.  The reward ablation follows the same
pattern: the direct paired contrast puts the graded reward $+.045$ above
binary top-one correctness at 8B (CI $[+.013,+.077]$), while at 1.7B the
two rewards are indistinguishable.  Graded credit for near misses helps
only at the larger scale.

The other two methods stay at or below their supervised references.  RFT amplifies
verifier-accepted traces the SFT student already samples; it matches SFT at
1.7B and loses ground at 8B under the matched recipe.  OPD receives no
captions, no labels, and no reward, only teacher distributions on
student-sampled SID trajectories, yet it still removes the interface
failures and recovers most of the teacher's gain without exceeding it
($73\%$ at 1.7B and $61\%$ at 8B).  Every trained model at both scales also
ranks above the caption-centroid ranker ($.184$), which neither Base
model reaches.  Complete paired contrasts and validity diagnostics are in
Appendix~\ref{app:anchor-results}.

\paragraph{History and slate size.}
\label{sec:scaling-results}
At the anchor setting, the SID prompt averages $845$ tokens against
$16{,}722$ for the item-matched text rendering; by $k{=}256$ the text
prompt exceeds $50$k tokens while the SID prompt stays under $2$k, a
$28\times$ gap (Figure~\ref{fig:scaling}a).  These counts use the
OpenOneRec tokenizer for every rendering; the providers' own tokenizers
count the anchor text prompt at $17.8$k to $31.2$k tokens, so the ratios
are conservative.  Over this range 8B RLVR
climbs with history, with most of the climb in place by $k{=}16$, below
the shortest history seen in training ($32$ items), and holds its plateau
through $k{=}256$, a significant paired gain over its own no-history setting
(Figure~\ref{fig:scaling}b).  The text curve comes from Gemini because the
text prompt nears the $32$k-token context of the OpenOneRec checkpoints at
$k{=}128$ and exceeds it at $k{=}256$ \citep{openonerec}.  The
Gemini reference gains nothing past the anchor setting: it peaks there
and drifts lower across the shaded region ($-.019$ by $k{=}256$, CI
$[-.050,+.011]$) while its prompt passes $50$k tokens.  The Base model
drifts with no sustained trend (Section~\ref{sec:zeroshot}), so the
response to history comes from post-training.  Histories are nested, so
each larger $k$ prepends the same user's earlier real interactions, shown
as SIDs to one system and as captions to the other, and 8B RLVR keeps
its gains from this added evidence at lengths where the text reader stops
improving.  One user shows the pattern (Appendix~\ref{app:qualitative},
Case~5).  The history returns again and again to chat-record screenshots
and emotional-confession videos, and the positive is another such video.
From the candidates alone, 8B RLVR leaves it out of the returned ten; at
$k{=}64$ and $k{=}256$ it ranks it second.  Gemini ranks it first at
$k{=}64$ but seventh at $k{=}256$, where its top pick switches to an
outfit-change video of a celebrity who dominates one stretch of the added
history.  Added
candidates lower every score, Random's included, since one positive now
competes with more negatives (Figure~\ref{fig:scaling}c).  8B RLVR
declines in parallel with the Gemini reference and ends level
with it at $N{=}100$, while its Base collapses toward Random.

\paragraph{The no-history margin.}
With no history at all, 8B RLVR beats Random by $+.079$
(CI $[+.039,+.118]$; Figure~\ref{fig:scaling}b, shaded setting).  A
$k{=}0$ prompt contains no history and no user identifier, so the margin
can only come from the candidate SIDs, whose embeddings are learned in
part from interaction sequences.  On the same slates rendered as captions,
the Gemini reference is statistically indistinguishable from Random.  8B
RLVR therefore uses item-level preference signal carried by
these embeddings, which a strong general reader does not recover from the
item text.

\paragraph{Rationale quality.}
\label{sec:rationale-quality}
Since 8B RLVR ranks on par with the text references, we
next compare its rationales with theirs.  For
$100$ anchor users we take the rationales that 8B RLVR and the
Gemini-3.6 reference actually generated during their
evaluations, verbatim, and have Claude-Opus-5 \citep{claudeopus5} compare
each pair aspect by aspect against the caption-rendered history and slate.  Each pair is judged
in both presentation orders, and a verdict counts only when both orders
agree.  The judge grades item-level claims, the user portrait, the
trade-off argument, coverage, and an overall verdict; coverage is also
counted directly from the text.  Appendix~\ref{app:rationale-judge} gives the protocol,
counts, and judge validity checks.

\begin{figure}[t]
  \centering
  \begin{minipage}[c]{0.55\textwidth}
    \includegraphics[width=\linewidth]{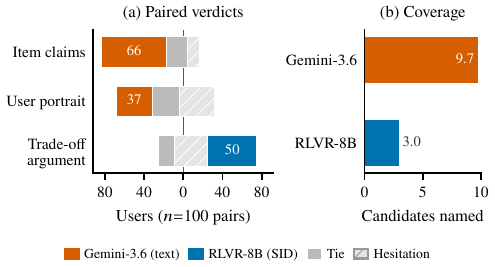}
  \end{minipage}\hfill
  \begin{minipage}[c]{0.42\textwidth}
    \caption{Paired rationale audit at the anchor setting ($100$ users; each pair
    judged in both presentation orders, and only order-consistent verdicts
    are decided).  (a)~Decided verdicts for three of the five judged
    aspects; hatched segments are hesitations, where the two presentation
    orders split between a win and a tie.  Judged coverage and the overall
    verdict, which tracks it, go to Gemini (Table~\ref{tab:rationale-judge}).  (b)~Mean number of the ten ranked candidates
    that the rationale names, counted by the fixed parser.}
    \label{fig:rationale-judge}
  \end{minipage}
\end{figure}

The verdicts split by aspect (Figure~\ref{fig:rationale-judge}).  RLVR
wins the trade-off aspect $50{:}0$ and discusses its candidates in the
order it ranks them (per-user Kendall $\tau{=}1.0$; $.983$ for Gemini,
which discusses more of them).  Its descriptions are weaker.  It names
about three of the ten ranked candidates where Gemini names nearly all
ten, and it loses item-level claims $0{:}66$ and the user portrait
$0{:}37$.  These errors are what the input predicts, since a model that
has never read a caption can state the user's interest correctly yet
attach it to an invented item.  In one pair (Appendix~\ref{app:qualitative},
Case~4), RLVR ranks the positive third and Gemini eighth, yet RLVR
justifies its top pick as a recap of a novel whose title appears nowhere
in the slate or the history.  The candidate is a recap of a different
novel, and the history does return repeatedly to novel recaps and
emotional stories; Gemini describes the same candidate correctly in one
line.  The judge is also stable: across
the $500$ paired aspect judgments, five per user pair, the decided
direction reverses once, and on
coverage it agrees with the programmatic count on every user where that
count has a winner.  Under SID-only input, ranking accuracy and
rationale reliability therefore diverge, and post-training taught the 8B
model to order the slate but not to describe what is in it.

\raggedbottom
\section{Conclusion}
\label{sec:conclusion}

We presented a controlled study of reason-then-rerank with SIDs as the
sole item representation, on \sysname{}, a testbed built for this
setting.  The existing SID-pretrained models fail at the interface: they
often return the input order unchanged, gain nothing from longer
histories, and do not improve when given readable captions.  Post-training
elicits the behavior.  SFT teaches the reason-then-rank interface at both
scales, and at 8B, RLVR with a graded ranking reward improves on SFT,
matches text-reading LLMs from prompts about $20\times$ shorter, ranks
above random without any history, and keeps its gains at long histories
where the text reference stops improving.  The rationale audit shows
the limit of these methods.  8B RLVR orders the slate well and argues
its trade-offs better than the text reference, but it misstates the
items it ranks.

These results show what standard post-training already gets out of the
SID tokens.  The tokens carry item-level preference signal that a strong
text reader does not recover from captions, since from the candidates
alone 8B RLVR beats Random and the caption reader does not.  They also
make long histories usable, because each added history item costs a few
SID tokens rather than a caption, and 8B RLVR keeps its gains at lengths
where the text reader stops improving.  The item descriptions are not yet
faithful, however: 8B RLVR states interests the history supports but
attaches them to items it misidentifies.  SFT imitates a teacher that
wrote from captions, and the verifiers score only the final ranking, so
post-training that also grades the reasoning is a natural next step for
reasoning recommenders.  Until then, a practical design is to rank from
SIDs, which keeps long histories cheap and uses behavioral signal, and
then write the explanation from retrieved metadata for the few items the
ranking places on top.  The testbed's text
rendering can also serve on its own as a public evaluation of how well general
LLMs model real user behavior.

\subsection*{AI Use Statement}

In this work, we used generative AI models as components of the
experimental method, as described in the paper: Gemini-3.6-Flash wrote the
hindsight rationales and rankings that form the SFT training data
(Section~\ref{sec:sft}, Appendix~\ref{app:sft}), Claude-Opus-5 served as the
rationale judge (Section~\ref{sec:rationale-quality},
Appendix~\ref{app:rationale-judge}), and GPT-5.6-Terra, Gemini-3.6-Flash,
and Claude-Sonnet-5 were evaluated as text-view reference systems
(Section~\ref{sec:zeroshot}).  We also used generative AI coding
assistants to help implement the experiment code, including the benchmark
builders, scoring and statistics scripts, post-training pipelines, and
plotting code.  Generative AI tools also helped draft and revise prose,
restructure \LaTeX{} source, and check formatting compliance.  The authors
reviewed all AI-assisted code, verified the scientific claims,
experimental choices, data handling, figures, citations, and final
manuscript, and take responsibility for all content of this work,
including anything produced with the aid of generative AI.

\subsection*{Ethics Statement}

The study uses the OpenOneRec research release of Kuaishou interaction
data under its access terms; we collect no new data, and users appear
only as opaque numeric identifiers.  Prompts expose item captions and
interaction histories, never user identifiers; caption text goes
to commercial LLM APIs only for the text references, the teacher
annotations, and the rationale judge.  These calls use the providers'
no-training, zero-retention settings.  Printed excerpts translate
user-facing text and redact personal names.  The testbed measures
engagement prediction for diagnostic purposes; deployment of its
rankings would require platform-side review of fairness and exposure
effects.

\subsection*{Reproducibility Statement}

Appendix~\ref{app:benchmark} specifies the testbed: eligibility
rules, the construction of users, slates, and negatives, the nested history
and slate interventions, and the split sizes.
Appendix~\ref{app:evaluation} gives the metrics, the two-stage inference
protocol, all decoding settings, the bootstrap procedure, and the exact
evaluation and extraction prompts.  Appendix~\ref{app:posttrain} gives, for
each post-training method, the data construction, objective, hyperparameters,
checkpoint-selection rule, and development trajectories, including the
exact SFT teacher prompt.  Appendix~\ref{app:results} reports the complete
anchor metrics and paired contrasts, and Appendix~\ref{app:artifacts}
describes the \anonword release of data builders, split manifests,
per-user result files, parsers, reward code, and run commands, together
with compute.  The \anonword code is linked in Section~\ref{sec:intro}.

\flushbottom
\bibliographystyle{iclr2027_conference}
\bibliography{rerankid-references}

\clearpage
\appendix
\counterwithin{table}{section}
\counterwithin{figure}{section}
\counterwithin{equation}{section}

\section*{Appendix Overview}

Appendices~\ref{app:benchmark} and~\ref{app:evaluation} specify the
benchmark and the evaluation protocol, including the exact prompts.
Appendix~\ref{app:posttrain} specifies the four post-training recipes.
Appendix~\ref{app:results} collects the additional results: the remaining
anchor metrics, paired contrasts, diagnostic breakdowns, robustness
audits, and qualitative cases.  Appendix~\ref{app:artifacts} describes the
released artifacts and compute.

\section{Benchmark Construction}
\label{app:benchmark}

\subsection{Task and Renderings}
\label{app:task}

Let $H=(h_1,\ldots,h_T)$ be a user's chronologically ordered history and
$C=(c_0,\ldots,c_{N-1})$ a candidate slate.  Given the last $k$
eligible history items and all $N$ candidates, a model returns a rationale and
an ordered JSON array of ten distinct candidate indices.  The primary
condition renders every item only as its three-codebook Semantic ID (SID).
The item-matched text and combined views change only the item rendering:
text replaces each SID with a dense caption, and the combined view shows
each item with both its SID and its caption.  Users,
items, labels, display order, and the index-valued action space remain fixed.
The phrase ``SID-only'' refers to the student's inference-time item view.  It
does not imply that post-training supervision contains no text.

\subsection{Users and Slates}
\label{app:users}

The source contains 162,074 users.  An evaluation user must have a renderable
positive and at least 256 history items that have both SID and caption views.
The development and test splits each contain 400 users, with 200 exact-one
grade-3 users and 200 exact-one grade-2 users.  A grade-3 positive is a like,
follow, or forward.  A grade-2 positive is a long-view event for a user with
no competing grade-3 event.  No user crosses the SFT, shared post-training,
benchmark-development, or benchmark-test partitions, and every history
temporally precedes its target.

For each evaluation user, a 100-item priority pool contains the positive,
clean exposure-hard negatives, and uniform-random catalog fill.  Additional
long-view items for grade-3 users are excluded rather than labeled negative.
Candidate sets are nested at $N\in\{10,20,30,50,100\}$.  Histories are nested
at $k\in\{0,4,8,16,32,64,128,256\}$ and retain the same recent unified items
under every rendering.  Display order is a deterministic function of user,
slate size, and shuffle replicate.  The anchor is $k=64$, $N=30$, and display
seed zero.  Table~\ref{tab:benchmark-settings} summarizes the setup.

\begin{table}[!b]
  \centering
  \small
  \setlength{\tabcolsep}{4.5pt}
  \begin{tabular}{@{}p{0.26\textwidth}p{0.66\textwidth}@{}}
    \toprule
    \textbf{Setting} & \textbf{Value} \\
    \midrule
    Platform / SID vocabulary & Kuaishou short-video interactions / OpenOneRec \\
    Development / test & 400 / 400 user-disjoint examples \\
    Positive strata per split & 200 grade-2 and 200 grade-3 users \\
    Labels per example & Exactly one behaviorally observed positive \\
    Minimum renderable history & 256 items with both SID and caption views \\
    History intervention & $k\in\{0,4,8,16,32,64,128,256\}$, nested recent subsequences \\
    Slate intervention & $N\in\{10,20,30,50,100\}$, nested from one 100-item pool \\
    Negative construction & Clean exposure-hard negatives followed by uniform-random catalog fill \\
    Output & Ten distinct in-range candidate indices, ordered by predicted relevance \\
    Anchor & $k=64$, $N=30$, display seed zero \\
    Test diagnostics & Mean 6.6 exposure-hard negatives; median unified history length 477 \\
    Coverage & 100\% SID and caption coverage in every evaluation slate \\
    \bottomrule
  \end{tabular}
  \caption{Benchmark construction.  One intervention axis varies at a
  time from the anchor setting while users, labels, and the other axis remain fixed.}
  \label{tab:benchmark-settings}
\end{table}

\subsection{Label Construction}
\label{app:labels}

Each example contains one observed positive, so NDCG@5, NDCG@10, Recall,
and MRR are all functions of that positive's returned rank, the standard
offline protocol for implicit feedback.  Two constructions keep the labels
clean.  Competing positives are excluded from the candidate pool rather
than treated as negatives (Section~\ref{app:users}), so an item the user
engaged with never enters a slate as a negative.  The highest-priority
negatives are exposure-hard: the platform showed the item and the user did
not engage, which is behavioral evidence against it.  Every system ranks
the same slates under the same labels, and the paper's claims rest
on paired per-user comparisons.  Comparisons among the candidates are also
examined as argued text by the rationale audit of
Section~\ref{app:rationale-judge}.

\subsection{Pretraining Exposure}
\label{app:pretraining-exposure}

The evaluation users come from the OpenOneRec public release, part of the
platform interaction data on which the released checkpoints are
pretrained ($162{,}074$ users in the dump we build from).  What the
backbone knows about an SID token is learned in this pretraining, which
includes interaction sequences involving its item; this
representation-level knowledge is the object of study.

Three structural facts separate this from example-level
memorization, the concern that a checkpoint recalls that a specific
history continues with a specific target.  First, no prompt in the
benchmark contains a user identifier, and the $k{=}0$ setting contains no
history either, so at $k{=}0$ a memorized user-item association has no
key to look up; the no-history margin is measured exactly where
this channel is closed, and only item-level information remains
usable.  Second, post-training cannot install test-set
memories, because its users are disjoint from the evaluation users.
Third, 8B RLVR's anchor accuracy is on par with the strongest text
reference, a model with no exposure to this platform reading the same
evidence as captions, so the accuracy level itself does not require a
memorization explanation.

The training tasks of the release also differ in kind from the task
studied here.  They include catalog-level next-item generation, pointwise
label prediction over a single candidate, and recommendation explanation,
but no task presents a candidate slate and asks for a reasoned ordering.
The reranking interface is therefore new to the checkpoints.

\subsection{Scope of the Reference Set}
\label{app:reference-scope}

Each reference anchors one quantity the study varies.  Random sets the
score scale of these slates, Caption shows what a non-generative text
baseline achieves on the same users, and the three API references show
what strong general text readers achieve under the same task wording and
answer format.  All of them are scored on the same slates as the systems
under study, so every main comparison stays paired.

\FloatBarrier
\section{Evaluation Protocol}
\label{app:evaluation}

\subsection{Metrics and Inference}
\label{app:metrics}

NDCG@5 is primary.  We additionally report NDCG@10, Recall@5, Recall@10, and
MRR.  A valid final answer contains exactly ten distinct integer indices in
range.  An example still invalid after the complete inference protocol
receives zero.  Interface diagnostics include final validity, the literal
identity ranking $[0,1,\ldots,9]$, the stage-1 reasoning-loop rate, and use of
the extraction fallback.  Valid-only averages are never substituted for the
end-to-end scores.

SID inference uses a fixed two-stage protocol.  Stage 1 permits free reasoning
with greedy decoding.  If it does not produce a valid list, stage 2 receives
up to the first 1,200 characters of that reasoning and uses structured
decoding to extract the list.  Both stages and every failure are retained.
The reported final validity is measured after this fixed fallback, whereas
the loop statistic describes stage 1.  Text API references use the same
task and index format in a single call, with no structured output and no
extraction stage.

Table~\ref{tab:extraction-accounting} accounts for the fallback at the
anchor setting.  Extraction\% is the share of users for whom stage 1
produced no valid list, so stage-1 validity is its complement.  The
stage-1-only column rescores each system with every extraction-salvaged
response set to zero, isolating what the fallback contributes to the
end-to-end score.  Before post-training the fallback matters:
it supplies $+.055$ of the 8B SID Base score and up to $+.065$ under the
combined rendering at 1.7B.  The dependence is heavier with no history:
at $k{=}0$ the stage is invoked for $47.0\%$ of users at 1.7B and
$19.5\%$ at 8B, and stage-1-only scoring would lower the $k{=}0$ SID
Base scores from $.119$ to $.053$ and from $.140$ to $.119$.  After
post-training every SFT, RFT, and RLVR model invokes it for
at most $1.0\%$ of users and gains at most $+.004$ from it.

\begin{table}[t]
  \centering
  \small
  \setlength{\tabcolsep}{5.0pt}
  \begin{tabular}{@{}llrrrr@{}}
    \toprule
    \textbf{System} & \textbf{View} & \textbf{Extraction\%} &
    \textbf{Final valid\%} & \textbf{NDCG@5} & \textbf{Stage-1 only} \\
    \midrule
    Base 1.7B & SID  & 17.5 & 100.0 & .114 & .101 \\
    Base 1.7B & Text & 47.8 &  99.5 & .115 & .071 \\
    Base 1.7B & Both & 43.5 &  99.8 & .139 & .074 \\
    Base 8B   & SID  & 37.5 & 100.0 & .168 & .113 \\
    Base 8B   & Text & 13.0 &  99.8 & .159 & .140 \\
    Base 8B   & Both &  6.3 & 100.0 & .166 & .154 \\
    \midrule
    SFT 1.7B  & SID  &  0.5 & 100.0 & .225 & .225 \\
    SFT 8B    & SID  &  0.5 & 100.0 & .237 & .234 \\
    RFT 1.7B  & SID  &  0.3 & 100.0 & .223 & .223 \\
    RFT 8B    & SID  &  0.0 & 100.0 & .211 & .211 \\
    RLVR (NDCG@5) 1.7B & SID & 0.3 & 100.0 & .231 & .231 \\
    RLVR (NDCG@5) 8B   & SID & 0.8 & 100.0 & .277 & .276 \\
    RLVR (Hit@1) 1.7B  & SID & 0.3 & 100.0 & .232 & .232 \\
    RLVR (Hit@1) 8B    & SID & 1.0 & 100.0 & .232 & .228 \\
    \bottomrule
  \end{tabular}
  \caption{Forced-extraction accounting at the anchor setting.  Stage-1
  validity is the complement of Extraction\%.  Stage-1 only rescores the
  system with extraction-salvaged responses set to zero.}
  \label{tab:extraction-accounting}
\end{table}

All main comparisons use the same 400 users.  The primary uncertainty
summary is a percentile 95\% confidence interval from 10,000 user-level paired
bootstrap resamples of the method-minus-initializer NDCG@5 difference.  Table~\ref{tab:evaluation-settings} lists the
inference settings.

\begin{table}[t]
  \centering
  \small
  \setlength{\tabcolsep}{4.0pt}
  \begin{tabular}{@{}p{0.26\textwidth}p{0.66\textwidth}@{}}
    \toprule
    \textbf{Setting} & \textbf{Value} \\
    \midrule
    SID models & OpenOneRec/OneRec-1.7B-pro and OpenOneRec/OneRec-8B-Pro, pinned released revisions \\
    Text references & Gemini-3.6-Flash and Claude-Sonnet-5 through Vertex AI; GPT-5.6-Terra through OpenRouter; retained finish reason \\
    Shared top-$K$ & 10 candidate indices \\
    SID stage 1 & Greedy decoding, 2,048-token response budget \\
    SID stage 2 & Structured array extraction, 256-token budget, 1,200-character reasoning cap \\
    Text API decoding & Gemini: visible rationale, medium thinking, provider-default temperature, 8,192-token cap; Claude and GPT: 32,768-token cap \\
    Primary metric & End-to-end NDCG@5, with outputs still invalid after extraction scored zero \\
    Additional metrics & NDCG@10, Recall@5/10, MRR \\
    Interface diagnostics & Final validity, stage-1 loop, extraction use, literal identity ranking \\
    Uncertainty & 10,000 user-paired bootstrap resamples; seed zero \\
    Calibration rankers & Seeded Random (seed 7); Caption uses Qwen3-Embedding-0.6B centroid similarity \\
    \bottomrule
  \end{tabular}
  \caption{Evaluation and inference settings.}
  \label{tab:evaluation-settings}
\end{table}

\subsection{Shared Evaluation Prompt}
\label{app:eval-prompt}

The only system-prompt substitution across renderings is the representation
noun: ``Semantic ID'', ``text description'', or ``Semantic ID and its text
description''.  The faithful English translation comes first, and the
original Chinese prompt used in all runs follows it.  Bracketed spans in the
translation and brace-delimited placeholders in the original mark
row-specific data fields, not extra instructions.  Natural rewordings of
this prompt are audited in Section~\ref{app:permutation}.

\begin{promptbox}{Shared evaluation prompt (English translation)}
SYSTEM: You are a recommendation reranking engine.  Given the known user
information, which may be empty, and a set of candidate items represented as
[REPRESENTATION], rank the candidates from most to least relevant according
to the user's interests.

\medskip
USER: [CHRONOLOGICAL USER HISTORY, OR AN EXPLICIT STATEMENT THAT NO HISTORY IS
AVAILABLE]

\medskip
Candidates (N items, numbered 0 through N-1):

[0] [RENDERED CANDIDATE 0]

\ldots

[N-1] [RENDERED CANDIDATE N-1]

\medskip
Briefly analyze the user's interests and how well the candidates match them.
Refer to items by candidate number, for example ``candidate 7''.  After the
reasoning ends, select the ten candidates that you consider most relevant and
order them from most to least relevant.  Output a JSON array containing
exactly ten distinct candidate indices in the range 0 through N-1, in the
form [<most relevant index>, <second most relevant index>, \ldots].
\end{promptbox}

\begin{promptbox}{Shared evaluation prompt (exact UTF-8 original)}
\begin{CJK}{UTF8}{gbsn}
SYSTEM: 你是一个推荐重排序引擎。给定用户的已知信息（可能为空）和一组用语义ID表示的候选物品，请根据用户兴趣把候选按相关性从高到低排序。

\medskip
USER: \{USER\_CONTEXT\}

\medskip
候选物品（共 30 个，编号 0 到 29）：

\{30 NUMBERED SID CANDIDATES\}

\medskip
请简明扼要地分析用户兴趣与各候选的匹配程度，分析时请用候选编号（如「候选7」）指代物品，然后在思考结束后，从这 30 个候选中挑出你认为最相关的 10 个，按相关性从高到低排列（最相关在最前）。输出一个 JSON 数组，恰好包含 10 个互不相同的候选编号（取值 0 到 29），形如 [<最相关的编号>, <次相关的编号>, …]。
\end{CJK}
\end{promptbox}

\noindent
The schematic output format contains no literal demonstration indices, which
prevents a weak model from copying an example ranking.  When stage 1 fails,
the extraction stage uses the system message ``You are a ranking-result
extractor: read a ranking analysis and output the final index ordering.''  Its
user message restates $N$ and the ten-index requirement, supplies the capped
stage-1 analysis, and requests only one JSON array with the most relevant
index first.  This extraction prompt is fixed across all SID models.

\begin{promptbox}{Extraction-stage prompt (exact UTF-8 original)}
\begin{CJK}{UTF8}{gbsn}
SYSTEM: 你是一个排序结果抽取器：读取一段排序分析，输出最终的编号排列。

\medskip
USER: 任务：从 30 个候选（编号 0 到 29）中选出最相关的 10 个，并从高到低排序。\\
以下是之前的分析过程：\\
\{STAGE-1 REASONING\}

\medskip
请只输出一个 JSON 数组，恰好包含 10 个互不相同的编号（取值 0 到 29），最相关在前。
\end{CJK}
\end{promptbox}

The reported Gemini-3.6 reference uses the same text-rendered task prompt with
an explicit request to place its visible analysis inside
\texttt{<think>...</think>} before the index array.  We disable response-schema
forcing so that this rationale is retained, use medium thinking with the
provider-default temperature, and allow 8,192 output tokens.  At the anchor setting,
399 of 400 outputs follow the output format and 399 contain a
nonempty visible rationale.

\FloatBarrier
\section{Post-Training Recipes}
\label{app:posttrain}

The four post-training methods share the SID-only task and evaluation benchmark.
SFT starts
from each Base model, whereas RFT and RLVR start from the
corresponding selected SFT checkpoint.  OPD starts from Base and queries a
frozen, scale-matched SFT teacher on student-generated SID trajectories.
Each method is defined by its initializer, supervision, and budget, and is
compared against its own initializer.

RFT, OPD, and RLVR use one shared 12,000-user post-training pool that is
user-disjoint from SFT, benchmark development, and benchmark test.  Their
common 6,000-prompt training set is balanced by positive stratum and
distributed as $k=32$: 582, $k=64$: 1,785, $k=128$: 1,820, $k=256$: 1,813 and
$N=10$: 580, $N=20$: 905, $N=30$: 2,457, $N=50$: 1,167, $N=100$: 891.  We
refer to this below as the shared post-training prompt set.  All checkpoint
selection uses benchmark-development NDCG@5 at the anchor setting; test data is never
consulted before the final evaluation.

\subsection{Supervised Fine-Tuning}
\label{app:sft}

The SFT source pool contains 12,000 unique users, with 6,000 examples from
each positive stratum and one slate per user.  We sample only long-history
conditions, using $k\in\{32,64,128,256\}$ with probabilities
$.10,.30,.30,.30$.  Candidate counts follow $N\in\{10,20,30,50,100\}$ with
probabilities $.10,.15,.40,.20,.15$.  The realized $k$ and $N$ counts in
Table~\ref{tab:sft-data} describe this 12,000-example source pool.  Quality
checks reject nine annotations, leaving 11,991 accepted examples.  The final
split contains 11,031 training and 960 development examples.

\begin{table}[!htb]
  \centering
  \small
  \setlength{\tabcolsep}{4.0pt}
  \begin{tabular}{@{}p{0.28\textwidth}p{0.64\textwidth}@{}}
    \toprule
    \textbf{Setting} & \textbf{Value} \\
    \midrule
    Source / quality-accepted examples & 12,000 / 11,991 unique users \\
    Source positive strata & Grade 2: 6,000; grade 3: 6,000 \\
    Accepted positive strata & Grade 2: 5,994; grade 3: 5,997 \\
    Train / development & 11,031 / 960 accepted examples, with zero overlap \\
    Quality rejections & 9 total: grounding 4; length 4; correctness 1 \\
    Source-pool history counts & $k=32$: 1,179; $64$: 3,648; $128$: 3,692; $256$: 3,481 \\
    Source-pool slate counts & $N=10$: 1,225; $20$: 1,761; $30$: 4,816; $50$: 2,385; $100$: 1,813 \\
    Teacher & Gemini-3.6-Flash, Vertex AI, medium thinking, provider-default temperature \\
    Teacher response budget & 8,192 tokens; one format-correction retry \\
    Teacher view / student view & Item-matched caption / SID-only \\
    Acceptance gates & Exact ten-index format, observed positive at rank one, explicit grounding to that candidate, nonempty rationale at most 600 characters \\
    Answer-reveal handling & Prohibited by the teacher prompt and audited \\
    Behavioral labels & One observed positive; negative-tail order supplied by the teacher \\
    \bottomrule
  \end{tabular}
  \caption{SFT annotation and data settings.}
  \label{tab:sft-data}
\end{table}

The caption teacher receives a privileged field containing the observed
positive index.  It writes a hindsight rationale and ranks that candidate
first.  The student then receives the item-matched SID-only prompt and
imitates the teacher response with ordinary conditional next-token
likelihood,
\begin{equation}
L_{\mathrm{SFT}}(\theta)=
-\sum_{(x,\,y^\ast)\in\mathcal{D}_{\mathrm{SFT}}}\;
\sum_{t=1}^{|y^\ast|}\log p_\theta\!\left(y^\ast_t\mid x,\,y^\ast_{<t}\right),
\label{eq:sft}
\end{equation}
where $x$ is the SID-only prompt, $y^\ast$ is the accepted teacher
response, rationale followed by the index array, and $\theta$ are the LoRA
parameters.  The teacher rationale is therefore a hindsight training trace.  The complete teacher instruction is
translated below and followed by the original Chinese prompt used in all runs.
Table~\ref{tab:sft-optimization} gives the optimization settings.

\begin{promptbox}{SFT hindsight-teacher prompt (English translation)}
SYSTEM: You are an assistant that simulates the internal reasoning process of
a recommendation reranking model.  You will see user context, numbered
candidates rendered as text descriptions, and one truly relevant candidate
index supplied only for internal construction.  Based on the user's history,
write a concise analysis that reads like an independent judgment and provide
a top-10 index list.  Never mention an internal reference, correct answer,
reference answer, being told, ground truth, or answer key.  Refer to items only
as ``candidate X''.  Do not copy candidate descriptions or enumerate every
candidate at length.

\medskip
USER: [TEXT-RENDERED USER CONTEXT]

\medskip
Candidates (N items, numbered 0 through N-1):

[0] [CANDIDATE 0 DESCRIPTION]

\ldots

[N-1] [CANDIDATE N-1 DESCRIPTION]

\medskip
Relevant candidate index supplied for internal supervision: [GOLD INDEX].
The other candidates have no observed positive feedback, but this does not
mean that they are semantically unrelated.  Still distinguish their degrees
of relevance from the user's history.

\medskip
Writing requirements:

1. In three to five concise sentences, summarize interests supported by the
history and explain why the first-ranked candidate matches.  Keep the language
proportional to the evidence.  Avoid exaggerated or absolute claims such as
``extremely'', ``precisely'', ``highly'', ``strongly'', ``perfectly'', or
``inevitably''.  When evidence is sparse, use restrained language such as
``may'' or ``to some extent''.

2. Consider both recent interests and interests in the longer history.  Only
when the history supports it, distinguish a currently prominent interest, a
stable interest that recurs over the longer history, and an incidental topic.
Do not invent preferences merely to cover both recent and long-term evidence,
or describe a rare topic as stable.  History items have no candidate numbers,
so do not invent labels such as ``history X''.

3. Explicitly mention the first-ranked item as ``candidate X'' and briefly
compare one or two alternatives to explain why the first choice fits better.
If one alternative is discussed, it must be second in the final array.  If two
are discussed, they must be second and third.  The judgments in the analysis
must agree with the final ranking.  Do not enumerate every candidate or copy
the full top-10 list into the analysis.

4. Judge all candidates by semantic fit to the complete user history.  Put the
observed relevant candidate first.  Rank every remaining candidate by its own
fit to the history, not merely by similarity to the first choice.  Before
ranking, check for major interests supported by multiple history events.  If a
candidate matches such an interest, place it appropriately even if its topic
differs from the first choice.  Do not allocate positions mechanically across
topics.

5. The response must begin with <think>.  Put a nonempty analysis between
<think> and </think>, then start a new line and output one JSON array.  The
array must contain exactly ten distinct indices in the range 0 through N-1.
The complete ranking may appear only once and only after </think>.  An array
without analysis, or before the analysis, is invalid.

\medskip
FORMAT-CORRECTION SUFFIX, USED ONCE IF REQUIRED: Independently redo the task.
The response must begin with <think>, end a nonempty analysis with </think>,
and then output exactly one JSON array containing ten distinct, in-range
candidate indices.  Do not add another tag, heading, or array.
\end{promptbox}

\begin{promptbox}{SFT hindsight-teacher prompt (exact UTF-8 original)}
\begin{CJK}{UTF8}{gbsn}
SYSTEM: 你是一个模拟推荐重排序模型内部推理过程的助手。你会看到用户上下文、若干个带编号的候选，候选用文字描述呈现，以及仅供内部使用的一个真实相关候选编号。请根据用户历史写出读起来像独立判断的简明分析，并给出 top-10 编号列表。不要提及内部参考、正确答案、参考答案、被告知、ground truth 或 answer key。分析时只用「候选X」指代物品，不要复述候选描述，也不要逐项长篇罗列。

\medskip
USER: \{CAPTION-RENDERED USER CONTEXT\}

\medskip
候选物品（共 30 个，编号 0 到 29）：

\{30 NUMBERED CAPTION CANDIDATES\}

\medskip
供内部构造监督使用的相关候选编号：[\{GOLD INDEX\}]。其余候选没有观察到正反馈，但这不代表它们语义上完全不相关；仍需根据用户历史区分其相关程度。

\medskip
写作要求：

1. 用简洁的 3-5 句话概括有历史证据支持的用户兴趣，并说明首选候选的匹配依据，不要写得过长。措辞应与历史证据的强度相称，避免使用「极其」「精准」「高度」「强烈」「完美」「必然」等夸张或绝对化判断；证据较少时使用「可能」或「一定程度」等克制措辞。

2. 同时考虑近期兴趣和更长历史中的兴趣，但只在历史证据支持时区分近期突出兴趣、长历史中反复出现的稳定兴趣和偶然主题。不要为了覆盖近期与长期而强行编造偏好，也不要把少量出现的主题描述成稳定兴趣。历史记录本身没有候选编号，不要编造「历史X」之类的历史编号。

3. 用「候选X」明确提到首选候选，并简要比较一到两个备选候选，说明首选为什么更合适。若比较一个，它必须是最终数组中的第二名；若比较两个，它们必须恰好是最终数组中的第二名和第三名。分析中的判断必须与最终排序一致。不要逐个候选罗列，也不要在分析中列出或复述完整 top-10。

4. 最后按与用户完整历史的语义匹配程度对所有候选进行判断。把相关候选排在第一位；除首选外，其余候选也应分别按照它们与完整用户历史的匹配程度排序，不要仅按照它们与首选候选的相似程度排序。排序前检查历史中有多条记录支持的主要兴趣：若候选中有相应物品，应按证据纳入合适位置，不要因为它与首选主题不同而忽略；也不要为不同兴趣机械分配名额。

5. 回复必须以 `<think>` 开头。先输出 `<think>` 和 `</think>` 包围的非空分析，再另起一行输出 JSON 数组。数组必须恰好包含 10 个互不相同且范围在 0 到 29 的编号，且完整排序只能在 `</think>` 后出现一次。只输出数组或把数组放在分析前均无效。

\medskip
FORMAT-CORRECTION SUFFIX, USED ONCE IF REQUIRED: 输出校正：请重新独立完成上述任务。回复必须严格以 `<think>` 开始，以 `</think>` 结束非空分析，然后只输出一次包含 10 个互不相同且范围合法的候选编号的 JSON 数组，不要添加其他标签、标题或数组。
\end{CJK}
\end{promptbox}

\begin{table}[H]
  \centering
  \small
  \setlength{\tabcolsep}{4.0pt}
  \begin{tabular}{@{}p{0.28\textwidth}p{0.30\textwidth}p{0.30\textwidth}@{}}
    \toprule
    \textbf{Setting} & \textbf{1.7B Pro} & \textbf{8B Pro} \\
    \midrule
    Initialization & Base model & Base model \\
    Adaptation & LoRA $r=16$, $\alpha=32$, all linear layers & Same \\
    Frozen weights & Base parameters, including SID embeddings & Same \\
    Template / cutoff & Qwen3 / 8,192 tokens & Qwen3 / 4,096 tokens \\
    Precision & bf16 & bf16 \\
    Effective batch & 32 & 128 \\
    Learning rate / schedule & $10^{-4}$ / cosine & $5\times10^{-5}$ / cosine \\
    Warmup & .08 & .08 \\
    Training epochs & 3 & 3 \\
    Checkpoint interval & 0.5 epoch & 0.5 epoch \\
    Selection criterion & Benchmark-dev NDCG@5 at $k=64,N=30$ & Same \\
    Selected epoch & 3.0 & 1.01 \\
    \bottomrule
  \end{tabular}
  \caption{SFT optimization settings.  Both scales use the same
  accepted examples (11,031 training, 960 development).  The 8B hyperparameters are selected by
  benchmark-development NDCG@5 before the test evaluation.}
  \label{tab:sft-optimization}
\end{table}

\FloatBarrier
\subsection{Rejection-Sampling Fine-Tuning}
\label{app:rft}

RFT is an offline self-training baseline initialized from SFT and trained on
the shared post-training prompt set.  For each SID-only prompt, the frozen
SFT model samples eight responses.  The model never sees the positive
label.  An offline verifier scans responses in generation order and retains
at most the first response that closes its reasoning, returns a valid
nonidentity list, and places the observed positive at rank one.  A response is rejected for an
invalid or unclosed answer, generation-length truncation, an empty rationale,
a positive below rank one, or identity collapse.  The resulting 1.7B dataset
contains 1,940 traces, corresponding to 32.33\% prompt coverage.  The matched
8B dataset contains 1,912 traces, or 31.87\% coverage.  Coverage is nearly
flat across history length and positive grade, but declines sharply as the
slate grows (Table~\ref{tab:rft-coverage}).

\begin{table}[!htb]
  \centering
  \small
  \setlength{\tabcolsep}{5.0pt}
  \begin{tabular}{@{}rrrrrr@{}}
    \toprule
    $N$ & \textbf{Source} & \textbf{1.7B retained} & \textbf{1.7B coverage} &
    \textbf{8B retained} & \textbf{8B coverage} \\
    \midrule
    10  & 580   & 360 & 62.07\% & 315 & 54.31\% \\
    20  & 905   & 373 & 41.22\% & 354 & 39.12\% \\
    30  & 2,457 & 826 & 33.62\% & 843 & 34.31\% \\
    50  & 1,167 & 267 & 22.88\% & 274 & 23.48\% \\
    100 & 891   & 114 & 12.79\% & 126 & 14.14\% \\
    \bottomrule
  \end{tabular}
  \caption{RFT retention by candidate count from the shared 6,000-prompt
  source population.  At most one response is retained per source prompt.}
  \label{tab:rft-coverage}
\end{table}

RFT reuses the shared evaluation prompt of Section~\ref{app:eval-prompt}; it
has no teacher prompt and no privileged label in the model input.  Continuing
the corresponding SFT adapter on the accepted responses is ordinary offline
likelihood training: the objective is the SFT loss of
Equation~\ref{eq:sft} with the accepted set in place of the teacher set,
where each accepted response $\hat y$ is the first of the eight samples
from the frozen $\pi_{\mathrm{SFT}}(\cdot\mid x)$ that passes the
verifier.  The sampling step is on the frozen initializer, but the
subsequent update is not online policy optimization.
Table~\ref{tab:rft-settings} gives the complete settings and
Table~\ref{tab:dev-curves} the development trajectories.

On development, no 1.7B RFT checkpoint exceeds the SFT initializer; the
best development checkpoint is step 62.  On test it reaches $.223$, compared with
$.225$ for SFT, a paired difference of $-.002$ with 95\% CI $[-.028,+.023]$.
At 8B the development curve does exceed the initializer and selects step 90,
but the selected checkpoint reaches $.211$ on test against $.237$ for SFT, a
paired difference of $-.026$ with 95\% CI $[-.047,-.005]$.

\begin{table}[!htb]
  \centering
  \small
  \setlength{\tabcolsep}{4.0pt}
  \begin{tabular}{@{}p{0.28\textwidth}p{0.64\textwidth}@{}}
    \toprule
    \textbf{Setting} & \textbf{Value} \\
    \midrule
    Models & OpenOneRec 1.7B Pro and 8B Pro \\
    Initialization & Corresponding dev-selected SFT checkpoint \\
    Source prompts & Shared post-training prompt set (6,000 SID-only examples) \\
    Samples per prompt & 8 \\
    Sampling & Temperature 1.0, top-$p$ .95, seed zero \\
    Response budget & 2,048 tokens \\
    Retention gate & First completed, valid, nonidentity response with positive at rank one; at most one per prompt \\
    Accepted traces & 1.7B: 1,940; 8B: 1,912 \\
    Update & Continue LoRA $r=16$, $\alpha=32$, all linear layers \\
    Cutoff / precision & 8,192 tokens / bf16 \\
    Effective batch & 32 \\
    Learning rate / schedule & $5\times10^{-5}$ / cosine, warmup .03 \\
    Training epochs & 2 \\
    Checkpoint interval & 0.5 epoch \\
    Selection criterion & Highest benchmark-dev NDCG@5 at $k=64,N=30$ \\
    Selected checkpoint & 1.7B: step 62; 8B: step 90 \\
    \bottomrule
  \end{tabular}
  \caption{RFT settings.  Both model scales use the same prompt keys,
  sampling hyperparameters, verifier, and update hyperparameters, while their
  sampled responses and accepted datasets remain separate.}
  \label{tab:rft-settings}
\end{table}

\FloatBarrier
\subsection{On-Policy Distillation}
\label{app:opd}

OPD tests whether a task-adapted model can provide dense supervision on the
states visited by a Base student.  At each scale, it uses a fresh LoRA on the
Base model as student and the frozen, merged SFT model at the same scale
as teacher.  Both receive the exact SID-only prompt of
Section~\ref{app:eval-prompt}.  The student samples the response, and at
response position $t$ the teacher receives the same prompt together with the
student prefix $y_{<t}$.  During OPD, neither model is directly supplied
with a caption, positive label, or answer hint, and no teacher-generated
trajectory is used.

For state $s_t=(x,y_{<t})$, let $V_{32}(s_t)$ be the teacher's 32
highest-probability tokens.  We renormalize the retained teacher mass,
\[
 \widetilde p_T(v\mid s_t)=
 \frac{p_T(v\mid s_t)}
 {\sum_{u\in V_{32}(s_t)}p_T(u\mid s_t)},\qquad
 v\in V_{32}(s_t),
\]
and minimize
\[
 L_t=\sum_{v\in V_{32}(s_t)}
 \widetilde p_T(v\mid s_t)
 \left[\log \widetilde p_T(v\mid s_t)-\log p_S(v\mid s_t)\right].
\]
The student probability remains normalized over its full vocabulary.  This
is a computational top-$32$ approximation to forward distillation, not a task
reward.  Loss is applied only to sampled response tokens, averaged within
each sequence, then across the minibatch.  Task reward, policy-gradient loss,
and a separate reference KL are disabled.  Table~\ref{tab:opd-settings} gives
the full specification and Table~\ref{tab:dev-curves} the development
trajectories.

Selecting the checkpoint with the highest dev score gives step 75 at 1.7B and step 50 at 8B.
On the test anchor, the selected 1.7B checkpoint reaches $.195$ against its
Base initializer at $.114$, a paired gain of $+.082$ with 95\% CI
$[+.044,+.120]$; it is valid for all anchor users, with no identity
permutation or stage-1 loop.  The selected 8B checkpoint reaches $.210$
against Base at $.168$, a paired gain of $+.042$ with 95\% CI
$[+.001,+.083]$, while its $-.026$ difference from SFT has 95\% CI
$[-.060,+.006]$.  It shows no identity permutation or stage-1 loop, and 395
of 400 outputs are valid after the fixed extraction fallback.

\begin{table}[!htb]
  \centering
  \small
  \setlength{\tabcolsep}{4.0pt}
  \begin{tabular}{@{}p{0.28\textwidth}p{0.64\textwidth}@{}}
    \toprule
    \textbf{Setting} & \textbf{Value} \\
    \midrule
    Models & OpenOneRec 1.7B Pro and 8B Pro \\
    Initialization / update & Base model / fresh LoRA $r=16$, $\alpha=32$, dropout 0, all linear layers \\
    Teacher & Frozen merged SFT model at the same scale \\
    Student and teacher input & Identical SID-only prompt and student-generated response prefix \\
    Training data & Shared post-training prompt set, label-free, one prompt per user \\
    Rollout & One response per prompt; temperature 1.0, top-$p$ 1.0 \\
    Prompt / response / model caps & 3,072 / 512 / 4,096 tokens \\
    Distillation & Renormalized teacher top-32 forward-KL approximation \\
    Loss aggregation & Response tokens only; sequence mean, then minibatch mean \\
    Other learning signals & No task reward, policy-gradient loss, or reference KL \\
    Teacher-query budget & 6,000 student trajectories, exactly one per training prompt \\
    Batch / horizon & 30 / 200 updates, one pass over the prompts \\
    Learning rate / warmup & $5\times10^{-5}$, constant after 6 warmup steps \\
    Checkpoints & Every 25 updates; all eight evaluated on 400 development users at each scale \\
    Runtime & Two A100 GPUs; verl 0.8.0; PyTorch 2.10.0; vLLM 0.19.1 \\
    \bottomrule
  \end{tabular}
  \caption{OPD specification.  Its feedback and initialization differ
  from the other methods, while its task, source prompt keys, and evaluation
  benchmark remain shared.}
  \label{tab:opd-settings}
\end{table}

\FloatBarrier
\subsection{Reinforcement Learning with Verifiable Rewards}
\label{app:rlvr}

RLVR starts from the selected SFT checkpoint and applies ordinary GRPO to a fresh
LoRA adapter, trained on the shared post-training prompt set.  The
benchmark development split at $k=64,N=30$ supplies 400 monitoring examples
but no policy-gradient updates.  Both runs complete 100 updates, with
checkpoints saved every 20 updates.

Let $r$ be the returned rank of the one observed positive.  The verifier uses
the benchmark's exact single-positive NDCG@5 reward:
\[
R(y)=
\begin{cases}
  1/\log_2(1+r), & \text{if $y$ is valid and } 1\le r\le5,\\
  0,             & \text{otherwise}.
\end{cases}
\]
A valid trajectory must close its \texttt{</think>} block and then emit only
one JSON array containing exactly ten distinct in-range integers.  A positive
outside the returned top ten, a rank below the NDCG cutoff, or any format
failure receives zero.  NDCG@10, Hit@1, identity, and parser fields are logged
only as diagnostics.  Identity receives no special penalty.  No rationale
content, rationale length, history-specific, or slate-specific shaping term is
added.

GRPO turns this reward into a group-relative policy gradient.  For each
prompt $x$, the model samples a group of $G{=}8$ responses, and each
response's advantage is its reward standardized within the group,
\begin{equation}
\hat A_i=\frac{R(y_i)-\operatorname{mean}\{R(y_j)\}_{j=1}^{G}}
{\operatorname{std}\{R(y_j)\}_{j=1}^{G}},
\end{equation}
so only within-group reward differences drive the update, with no learned
value model.  Every token of response $y_i$ shares the advantage
$\hat A_i$, and the update maximizes the clipped surrogate
\begin{equation}
J(\theta)=\mathbb{E}_{i,t}\!\left[
\min\!\big(\rho_{i,t}\hat A_i,\;
\operatorname{clip}(\rho_{i,t},1-\varepsilon,1+\varepsilon)\,\hat A_i\big)
\right]
-\beta\,D_{\mathrm{KL}}\!\big(p_\theta\,\|\,p_{\mathrm{SFT}}\big),
\end{equation}
where the expectation averages over sampled response tokens, $\rho_{i,t}$
is the token-level probability ratio between the current model and the
model that sampled the group, $p_{\mathrm{SFT}}$ is the merged SFT
initializer, and $\beta{=}.001$ (Table~\ref{tab:rlvr-settings}).  The clip
range and loss aggregation follow the verl 0.8.0 defaults recorded with
the run configurations.

At both scales, a controlled reward ablation keeps the corresponding SFT
initializer and replaces the graded reward with
\[
R_{\mathrm{top1}}(y)=
\begin{cases}
1, & \text{if $y$ is valid and } r=1,\\
0, & \text{otherwise}.
\end{cases}
\]
All other data, rollout, optimizer, checkpoint, and development-selection
settings are unchanged.  RLVR reuses the shared SID prompt of
Section~\ref{app:eval-prompt} and has no teacher prompt.
Table~\ref{tab:rlvr-settings} gives the full settings and
Table~\ref{tab:dev-curves} the development trajectories.

\begin{table}[t]
  \centering
  \small
  \setlength{\tabcolsep}{4.0pt}
  \begin{tabular}{@{}p{0.28\textwidth}p{0.64\textwidth}@{}}
    \toprule
    \textbf{Setting} & \textbf{Value} \\
    \midrule
    Models & OpenOneRec 1.7B Pro and 8B Pro \\
    Primary initialization & Corresponding dev-selected SFT checkpoint, merged into the base \\
    Trained weights & Fresh LoRA $r=16$, $\alpha=32$, all linear layers \\
    Training / monitor data & Shared post-training prompt set / 400 benchmark-dev users \\
    Rollouts per prompt & 8 \\
    Sampling & Temperature 1.0, top-$p$ .95 \\
    Prompt / response / model caps & 6,144 / 2,048 / 8,192 tokens \\
    Training duration & 100 optimizer steps at each scale \\
    Global prompt batch / PPO mini batch & 60 / 30 \\
    PPO epochs & 1 \\
    Learning rate / warmup & $10^{-5}$ / .03 \\
    KL coefficient & .001 against the merged SFT initializer \\
    Evaluated checkpoints & Steps 20, 40, 60, 80, 100 at each scale \\
    Selection criterion & Highest official two-stage benchmark-dev NDCG@5 among completed RLVR checkpoints; earlier checkpoint breaks ties \\
    Selected checkpoint & 1.7B: step 60; 8B: step 100 \\
    Reward ablation & Binary top-one reward; SFT initialization; 1.7B selects step 60 and 8B selects step 80 \\
    \bottomrule
  \end{tabular}
  \caption{GRPO RLVR settings.  Both scales share the data, reward, rollout,
  and per-update optimization settings.}
  \label{tab:rlvr-settings}
\end{table}

\begin{table}[t]
  \centering
  \small
  \setlength{\tabcolsep}{4.5pt}
  \begin{tabular}{@{}llrrrrrrrrr@{}}
    \toprule
    \textbf{Method} & \textbf{Scale} & \textbf{Init} &
    \multicolumn{8}{c}{\textbf{Checkpoint NDCG@5 on development}} \\
    \midrule
    RFT (steps 31 to 122 / 30 to 120) & 1.7B & .220 & .217 & \textbf{.217} & .209 & .217 & & & \\
     & 8B & .217 & .222 & .250 & \textbf{.254} & .253 & & & \\
    \addlinespace
    OPD (steps 25 to 200) & 1.7B & -- & .129 & .139 & \textbf{.211} & .197 & .206 & .185 & .203 & .202 \\
     & 8B & -- & .167 & \textbf{.266} & .227 & .227 & .246 & .243 & .228 & .234 \\
    \addlinespace
    RLVR (steps 20 to 100) & 1.7B & .218 & .220 & .223 & \textbf{.233} & .233 & .225 & & \\
     & 8B & .230 & .230 & .224 & .244 & .248 & \textbf{.301} & & \\
    \addlinespace
    RLVR top-one (steps 20 to 100) & 1.7B & .211 & .211 & .211 & \textbf{.218} & .213 & .196 & & \\
     & 8B & .214 & .230 & .230 & .235 & \textbf{.259} & .251 & & \\
    \bottomrule
  \end{tabular}
  \caption{Development-selection trajectories at the anchor setting.  Checkpoints are
  listed in training order; the selected checkpoint is bold.  RFT evaluates
  four checkpoints (1.7B: steps 31, 62, 93, 122; 8B: steps 30, 60, 90, 120),
  OPD eight (every 25 updates), and RLVR five (every 20 updates).  For the
  1.7B RFT row the bold checkpoint is the highest-scoring RFT checkpoint even
  though it does not exceed the initializer.  Init is the initializer's dev
  score under each method's evaluation harness; OPD initializes from the Base model, whose
  anchor scores appear in Table~\ref{tab:main-anchor}.}
  \label{tab:dev-curves}
\end{table}

On test, the selected 1.7B RLVR checkpoint reaches $.231$ against $.225$ for
SFT, a paired difference of $+.006$ with 95\% CI $[-.015,+.026]$.  The
selected 8B checkpoint reaches $.277$ against the SFT result at
$.237$, a paired gain of $+.040$ with 95\% CI $[+.006,+.076]$; relative to
Base, its cumulative gain is $+.109$ with CI $[+.069,+.149]$.  All reported
RLVR outputs are valid, with no identity permutation or stage-1 loop.

The reward ablations are also fully valid.  At 1.7B, the top-one model
reaches $.232$ on test, a $+.007$ difference from SFT with 95\% CI
$[-.014,+.028]$ and a $+.002$ difference from the graded-reward model with
CI $[-.018,+.022]$.  At 8B, the top-one model reaches $.232$, a $-.004$
difference from SFT with 95\% CI $[-.025,+.016]$, while its $-.045$
difference from the graded-reward model has CI $[-.077,-.013]$.  Binary and
graded rewards are therefore not significantly different at 1.7B, while
at 8B the graded reward's gain is significant.

\FloatBarrier
\section{Additional Results}
\label{app:results}

This section collects the results that support the main text but do not
fit there: the remaining anchor metrics and paired contrasts, the
rendering, token, and stratum diagnostics, the display-order,
prompt-wording, and rationale-quality audits, and qualitative cases.

\subsection{Anchor Metrics and Paired Contrasts}
\label{app:anchor-results}

Table~\ref{tab:main-anchor} already reports the full anchor metric suite for the
calibration rankers, the text references, and the existing models under
all three renderings, together with NDCG@5 and the interface diagnostics for
every post-trained model (Table~\ref{tab:posttrain-anchor}).  Table~\ref{tab:full-anchor} therefore adds only
what is missing there: the remaining ranking metrics for the post-trained
models.  Table~\ref{tab:paired-comparisons} gives the full set of paired
contrasts.  Beyond the initializer intervals of Table~\ref{tab:posttrain-anchor},
only two kinds of added contrast are significant: the primary RLVR models' gains
over their Base models, at both scales, and the 8B graded-reward gain over
the binary reward.  Every added comparison against a text reference or
across scales stays within noise.
One further contrast compares the two RLVR rewards directly
rather than through their shared initializer: the graded NDCG@5 reward
finishes $+.045$ above binary top-one correctness at 8B
(user-paired bootstrap CI $[+.013,+.077]$), while at 1.7B the two rewards
are indistinguishable ($-.002$, CI $[-.022,+.018]$).

\begin{table}[t]
  \centering
  \small
  \setlength{\tabcolsep}{4.5pt}
  \begin{tabular}{@{}llllccccc@{}}
    \toprule
    \textbf{Method} & \textbf{Scale} &
    \textbf{Initializer} & \textbf{Supervision} &
    \textbf{N@5} & \textbf{N@10} & \textbf{R@5} & \textbf{R@10} &
    \textbf{MRR} \\
    \midrule
    SFT & 1.7B Pro & Base & Teacher traces & .225 & .290 & .358 & .560 & .208 \\
    OPD & 1.7B Pro & Base & Teacher distributions & .195 & .242 & .310 & .455 & .178 \\
    RFT & 1.7B Pro & SFT & Filtered rollouts & .223 & .290 & .358 & .568 & .206 \\
    RLVR & 1.7B Pro & SFT & NDCG@5 reward & .231 & .298 & .353 & .563 & .218 \\
    RLVR & 1.7B Pro & SFT & Hit@1 reward & .232 & .298 & .358 & .565 & .217 \\
    \midrule
    SFT & 8B Pro & Base & Teacher traces & .237 & .302 & .360 & .565 & .223 \\
    OPD & 8B Pro & Base & Teacher distributions & .210 & .264 & .318 & .485 & .197 \\
    RFT & 8B Pro & SFT & Filtered rollouts & .211 & .274 & .338 & .535 & .195 \\
    RLVR & 8B Pro & SFT & NDCG@5 reward & .277 & .347 & .425 & .640 & .257 \\
    RLVR & 8B Pro & SFT & Hit@1 reward & .232 & .304 & .353 & .573 & .222 \\
    \bottomrule
  \end{tabular}
  \caption{Ranking metrics for the post-trained models at the anchor setting
  ($k=64$, $N=30$, display seed zero, 400 test users).  N@$j$ and R@$j$
  denote NDCG@$j$ and Recall@$j$.  Anchors for Random, Caption, the text
  references, and the existing models appear in Table~\ref{tab:main-anchor}.  All
  rows are SID-only at inference.  Every row is fully valid with no identity
  permutation or stage-1 loop, except that 8B OPD yields a valid ranking for
  395 of 400 outputs.  Supervision terms are defined in Section~\ref{sec:controls}.  Repeated RLVR
  rows are controlled reward ablations.}
  \label{tab:full-anchor}
\end{table}

\begin{table}[t]
  \centering
  \small
  \setlength{\tabcolsep}{5.0pt}
  \begin{tabular}{@{}llrrrr@{}}
    \toprule
    \textbf{A} & \textbf{B} & \textbf{Mean A} & \textbf{Mean B} &
    $\boldsymbol{\Delta}$\textbf{(A-B)} & \textbf{95\% paired CI} \\
    \midrule
    SFT 1.7B & Base 1.7B & .225 & .114 & +.111 & $[+.071,+.150]$ \\
    SFT 8B & Base 8B & .237 & .168 & +.068 & $[+.028,+.109]$ \\
    GPT-5.6-Terra text & Gemini text & .256 & .247 & +.009 & $[-.015,+.032]$ \\
    GPT-5.6-Terra text & SFT 1.7B & .256 & .225 & +.031 & $[-.007,+.068]$ \\
    SFT 1.7B & Gemini text & .225 & .247 & -.022 & $[-.061,+.016]$ \\
    SFT 8B & Gemini text & .237 & .247 & -.011 & $[-.049,+.026]$ \\
    SFT 8B & SFT 1.7B & .237 & .225 & +.012 & $[-.020,+.043]$ \\
    OPD 1.7B & Base 1.7B & .195 & .114 & +.082 & $[+.044,+.120]$ \\
    OPD 1.7B & SFT 1.7B & .195 & .225 & -.030 & $[-.065,+.004]$ \\
    OPD 8B & Base 8B & .210 & .168 & +.042 & $[+.001,+.083]$ \\
    OPD 8B & SFT 8B & .210 & .237 & -.026 & $[-.060,+.006]$ \\
    RFT 1.7B & SFT 1.7B & .223 & .225 & -.002 & $[-.028,+.023]$ \\
    RFT 8B & SFT 8B & .211 & .237 & -.026 & $[-.047,-.005]$ \\
    RLVR 1.7B & SFT 1.7B & .231 & .225 & +.006 & $[-.015,+.026]$ \\
    RLVR Top-1 1.7B & SFT 1.7B & .232 & .225 & +.007 & $[-.014,+.028]$ \\
    RLVR Top-1 1.7B & RLVR 1.7B & .232 & .231 & +.002 & $[-.018,+.022]$ \\
    RLVR 8B & SFT 8B & .277 & .237 & +.040 & $[+.006,+.076]$ \\
    RLVR Top-1 8B & SFT 8B & .232 & .237 & -.004 & $[-.025,+.016]$ \\
    RLVR Top-1 8B & RLVR 8B & .232 & .277 & -.045 & $[-.077,-.013]$ \\
    RLVR 1.7B & Base 1.7B & .231 & .114 & +.117 & $[+.075,+.159]$ \\
    RLVR 8B & Base 8B & .277 & .168 & +.109 & $[+.069,+.149]$ \\
    RLVR 8B & Gemini text & .277 & .247 & +.029 & $[-.006,+.065]$ \\
    RLVR 8B & GPT-5.6-Terra text & .277 & .256 & +.021 & $[-.013,+.054]$ \\
    \bottomrule
  \end{tabular}
  \caption{User-paired NDCG@5 comparisons at the anchor setting.  Comparisons with
  text-view LLMs are diagnostic because model family and item representation
  both differ.}
  \label{tab:paired-comparisons}
\end{table}

\FloatBarrier
\subsection{Rendering, Tokens, and Engagement Strata}
\label{app:diagnostic-results}

The matched-view controls, which cross backbone scale with the three
item-matched input views at the anchor setting, appear in Table~\ref{tab:main-anchor}.  The only
diagnostic not reported there is final two-stage validity: the SID view
yields a valid list for every anchor example at both scales, and the text and combined
views remain between 99.5\% and 100\%.
Table~\ref{tab:token-economy} reports the matched prompt lengths.

\begin{table}[t]
  \centering
  \small
  \setlength{\tabcolsep}{5.0pt}
  \begin{tabular}{@{}lrrr@{}}
    \toprule
    \textbf{View} & $k=64$ & $k=128$ & $k=256$ \\
    \midrule
    SID & 845 & 1,165 & 1,805 \\
    Text & 16,722 & 27,899 & 50,412 \\
    Both & 17,255 & 28,752 & 51,905 \\
    \bottomrule
  \end{tabular}
  \caption{Mean input tokens under the OpenOneRec tokenizer at $N=30$.}
  \label{tab:token-economy}
\end{table}

The OpenOneRec tokenizer is one common yardstick across renderings, not
the billing count of any commercial API.  Measured from the recorded API
usage metadata, the mean anchor text prompt is $17{,}815$ tokens for
Gemini-3.6-Flash, $21{,}317$ for GPT-5.6-Terra, and $31{,}209$ for
Claude-Sonnet-5, each at or above the $16{,}722$ of
Table~\ref{tab:token-economy}, so the compactness ratios of Section~\ref{sec:scaling-results} are
conservative under the providers' own tokenizers.

The balanced SFT source pool yields a near-balanced accepted set (5,994
grade-2 and 5,997 grade-3 examples).  Table~\ref{tab:sft-strata} breaks
the SFT gain down by stratum: at both scales it is larger for grade-3
positives, the explicit likes, follows, and forwards.

\begin{table}[t]
  \centering
  \small
  \setlength{\tabcolsep}{5.0pt}
  \begin{tabular}{@{}lrrr@{}}
    \toprule
    \textbf{Scale / stratum} & \textbf{Base} & \textbf{SFT} &
    $\boldsymbol{\Delta}$ \textbf{[95\% CI]} \\
    \midrule
    1.7B / grade 2 & .132 & .206 & $+.074$ $[+.018,+.131]$ \\
    1.7B / grade 3 & .095 & .244 & $+.149$ $[+.093,+.204]$ \\
    8B / grade 2 & .186 & .223 & $+.037$ $[-.018,+.092]$ \\
    8B / grade 3 & .150 & .250 & $+.100$ $[+.040,+.160]$ \\
    \bottomrule
  \end{tabular}
  \caption{Anchor NDCG@5 by positive-engagement stratum.}
  \label{tab:sft-strata}
\end{table}

\FloatBarrier
\subsection{Display-Order and Prompt-Wording Stability}
\label{app:permutation}

Two parallel audits rescore the Gemini-3.6 text reference at the anchor setting,
one under new candidate display orders and one under new prompt wordings.
Both are judged against the same baseline: rerunning an unchanged
condition.  Because the reference decodes at the provider-default
temperature, an identical rerun already moves mean NDCG@5 by up to
$.014$, and the per-user rankings of two identical runs agree at Kendall
$\tau\approx.54$.

The display-order audit rescores the anchor setting under the two reserved display
orders and leaves the reference essentially unchanged: NDCG@5 is $.247$,
$.234$, and $.245$ across the three orders, and the original order differs
from the permutation average by $+.005$ with paired 95\% CI
$[-.007,+.018]$.  We had planned to switch to permutation-averaged
scores if this difference were significant; it is not, so all results keep
the original order.  In each of these orders, the
top-ranked choice is statistically indistinguishable from uniform over the
$30$ input positions ($\chi^2(29)\leq 27.7$ against a $42.6$ critical
value), so scores reflect item content rather than display position.  A
fourth order, scored together with the wording audit, leaves the original
order $+.0065$ above the four-order average.  The order sensitivity of each SID system is
reported separately through its identity-permutation rate.

The prompt-wording audit rescores the anchor setting under three natural
rewordings of the original prompt's task statement.  Each keeps the output
format, the history and candidate blocks, and the ``candidate $X$''
convention, so the audit varies phrasing rather than the task.  No
rewording moves the anchor score out of the rerun range: the paired differences
from the original prompt are $+.009$, $-.011$, and $-.002$, every 95\% CI
straddles zero, and the rate of outputs that follow the format is unchanged at
$399$ to $400$ of $400$.  The rescored runs and both audit reports
are included in the released artifacts.

\subsection{Rationale-Quality Judge}
\label{app:rationale-judge}

This audit compares rationale quality, not ranking quality, for the two
strongest rationale-emitting systems at the anchor setting: the 8B RLVR (NDCG@5
reward) model on the SID view and the Gemini-3.6-Flash text reference.
Both rationales are the reasoning blocks the two systems emitted during
their anchor evaluations, taken verbatim; nothing is regenerated,
because a regenerated rationale would describe a system the paper never
reports.

\paragraph{Protocol.}
For $100$ seeded test users, the judge (Claude-Opus-5, API model
\texttt{claude-opus-5} served through Vertex, medium reasoning
effort) receives the caption-rendered history and candidate slate, both
anonymized rationales, and both emitted rankings, and returns one verdict
(A, B, or tie) per aspect.  It is instructed not to infer the observed
positive, not to compare the rankings themselves, and not to reward length
or breadth outside the coverage aspect.  Every pair is judged twice, once
in each presentation order.  A verdict is decided only when the two orders
agree; cross-order disagreements are reported separately and never split.
Across the $786$ non-tie verdicts, the judge chose the first-presented
rationale $48.98\%$ of the time.  Order-consistent pairs contribute one
first- and one second-presented choice by construction, so the
informative subset is the $86$ non-tie verdicts from pairs that split
across orders: the first-presented rationale wins $35$ of them (sign test
$p{=}.11$), so presentation order shows no measurable preference.  Item-level and user-level accuracy
are kept as separate aspects: naming the wrong item is the expected
failure of a model that has never read a caption, whereas the user
portrait summarizes the history as a whole.  Averaging them would hide
this distinction.

\begin{table}[t]
  \centering
  \small
  \setlength{\tabcolsep}{4.5pt}
  \begin{tabular}{@{}lccccc@{}}
    \toprule
    \textbf{Aspect} & \textbf{Gemini} & \textbf{RLVR} & \textbf{Tie} &
    \textbf{Hesit.} & \textbf{Flip} \\
    \midrule
    Item-claim accuracy   & \textbf{66} & 0 & 21 & 12 & 1 \\
    User-model accuracy   & \textbf{37} & 0 & 28 & 35 & 0 \\
    Trade-off argument    & 0 & \textbf{50} & 16 & 34 & 0 \\
    Coverage (judged)     & \textbf{100} & 0 & 0 & 0 & 0 \\
    Overall               & \textbf{97} & 0 & 0 & 3 & 0 \\
    \bottomrule
  \end{tabular}
  \caption{Paired verdicts over $100$ users.  Gemini and RLVR columns count
  pairs decided for that system in both presentation orders; Hesit.\ counts
  cross-order hesitations between a win and a tie; Flip counts pairs where
  the decided direction reversed across orders.  Every one-sided sign test
  rejects at $p<10^{-10}$.  The judged coverage row and the overall verdict
  are reported for completeness: coverage is also measured programmatically
  below, and the overall verdict tracks it.}
  \label{tab:rationale-judge}
\end{table}

\paragraph{Results.}
Table~\ref{tab:rationale-judge} gives the counts behind
Figure~\ref{fig:rationale-judge}.  In $500$ paired aspect judgments the decided direction reversed
once; every other cross-order disagreement is a hesitation between a win
and a tie, so the per-aspect direction is stable even where hesitation is
common.  Two quantities are computed directly from the
rationale text rather than judged, using the fixed parser, which reads
each explicit candidate reference.  Coverage is the share of the ten
ranked candidates that the rationale names at least once: $.971$ for
Gemini ($9.7$ of the ten) against $.297$ for RLVR ($3.0$).  Fidelity is
the per-user Kendall $\tau$ between the order of first mention and the
emitted ranking, restricted to candidates that are both mentioned and
ranked, and averaged over users with at least two such candidates, which holds for
all $100$ users of both systems: $.983$ for Gemini and $1.0$ for RLVR.  Unmentioned candidates are
excluded rather than tied at the bottom, so that coverage does not leak
into the fidelity number.

\paragraph{Judge validity.}
We run two independent checks.  First, coverage has a programmatic proxy, the
count of named candidates, so the judged coverage row can itself be
scored: the computed winner is Gemini
for $98$ of $100$ users with $2$ exact ties, and the judge agreed on all
$98$ with zero disagreements.  Second, controlled corruption: for three
users we re-ran the protocol after corrupting the Gemini rationale in three
ways, swapping two candidate indices inside the text (falsifying item
claims while preserving length, style, and the discussed candidate set),
substituting another user's rationale, and truncating the rationale to its
top-one justification.  Aspects targeted by a corruption stopped being won
by the corrupted rationale in all eight applicable cases, and aspects that should
be unaffected held in three of four; the single miss became a cross-order
hesitation, not a reversed verdict.  The index swap never moved coverage,
as designed.

\paragraph{Systems compared.}
The audit pairs the strongest SID model with the text reference, the
comparison the main results draw.  Counting reasoning blocks that are
nonempty and terminated, every post-trained model completes a rationale
for all $400$ anchor users, and Gemini writes a nonempty visible rationale
for $399$.  The base models are less reliable: base-1.7B completes $50$
of $400$, mostly closing an empty block instead, and base-8B completes
$236$, with most of the rest transcribing prompt SID tokens until the
stage-1 token cap.  Judge
outputs, corrupted variants, and the control report are included in the
released artifacts.

\subsection{Qualitative Cases}
\label{app:qualitative}

Every case below is quoted from the released result files of the final runs;
Case 3 additionally quotes one example from the released SFT training
split.  Nothing is regenerated or edited.
Each case names the anonymized user row of the released build, the system,
and the setting; every evaluation shown uses the evaluation prompt.  Responses
and captions are Chinese.  Excerpts appear here in English translation with
whitespace compressed, SID tokens verbatim, and personal names replaced by
bracketed placeholders; the untranslated originals are in the released
files.

\paragraph{Case 1: identity collapse.}
User row 67158, 8B Base, SID view, anchor setting.  Stage 1 terminates
after $46$ output tokens with an empty reasoning block and the literal
identity ranking:
\begin{promptbox}{Case 1: the complete stage-1 response}
<think>

</think>[0, 1, 2, 3, 4, 5, 6, 7, 8, 9]
\end{promptbox}
\noindent
The response is format-valid, so no fallback triggers, and the observed
positive, at display position $29$, is dropped entirely.  Identity\%
counts every such input-order ranking: $82$ of $400$ anchor users for
this model, $75$ of which come from forced extraction after a loop like
Case~2, and $169$ for base-1.7B, $138$ of which come from an empty block
like this one.

\paragraph{Case 2: reasoning loop.}
User row 1671, same run.  Stage 1 opens with two planning sentences and then
transcribes the prompt's history SID groups, in prompt order, until the
$2{,}048$-token stage-1 cap; the reasoning block never closes:
\begin{promptbox}{Case 2: stage-1 excerpt (translated; SID tokens verbatim)}
<think> Okay, I now need to analyze the user's viewing history and the
candidate items and find the ten most relevant candidates.  First, I should
look carefully at the user's viewing history and the characteristics of
each candidate item.

The videos the user watched include:
<|sid\_begin|><s\_a\_2890><s\_b\_5369><s\_c\_3086><|sid\_end|>,
<|sid\_begin|><s\_a\_2180><s\_b\_1625><s\_c\_6870><|sid\_end|>, [the
transcription continues to the token cap]
\end{promptbox}
\noindent
All $149$ looping anchor users of the 8B Base stop at exactly this cap.
Forced extraction then returns the identity list for this user, and the
positive, at display position $26$, is again lost.

\paragraph{Case 3: the teacher demonstration and the student's view.}
The first example of the released SFT training split
(Section~\ref{app:sft}), a $64$-item history with a $20$-candidate training
slate.  The caption-reading teacher grounds its picks in readable content:
\begin{promptbox}{Case 3: teacher rationale excerpt (translated)}
The user's history shows a multi-dimensional distribution of interests.
Over the long term, the user pays fairly stable attention to rural daily
life, traditional home cooking, and warm family interaction; the recent
history leans more toward fashion and outfits, makeup transformations,
celebrity and appearance content, and emotionally resonant campus-life
stories. [...] Candidate 3 shows rural life and kitchen cooking, fitting
the user's long-standing preference for pastoral food content. [...]
\end{promptbox}
\noindent
The item-matched student instruction renders every one of the same items as
an SID group alone; its first history item is
\texttt{<|sid\_begin|><s\_a\_946><s\_b\_6523><s\_c\_444><|sid\_end|>}, and
no caption text appears anywhere in it.  The student imitates reasoning
grounded in text it cannot read, which is why instance-level item claims
are the expected failure of the post-trained models
(Section~\ref{app:rationale-judge}).

\paragraph{Case 4: what the RLVR model argues, and what it gets wrong.}
User row 88506, anchor setting: the selected 8B RLVR model against the
Gemini-3.6 text reference, a pair judged in
Section~\ref{app:rationale-judge}.  RLVR returns the positive at rank $3$;
Gemini returns it at rank $8$.  The RLVR reasoning block spends its $192$
tokens on exactly three candidates, and its narrative order matches its
returned ranking:
\begin{promptbox}{Case 4: the two rationales on the same slate (translated
excerpts)}
8B RLVR: ``Candidate 27 tells the plot and emotional entanglements of the
novel `Sisyphe Bookstore', highly matching the novel-recap and
emotional-story themes that appear repeatedly in the user's history. [...]
Candidate 12 leans toward campus youth and selfies, and candidate 8 toward
internet hot topics and comedic banter [...] slightly inferior to candidate
27 in thematic fit.''

True caption of candidate 27: ``A detailed interpretation of the
characters' emotional entanglements and complex relationships in the novel
`No Longer Devoted', finally revealing the story's ending.''

Gemini-3.6, same candidate: ``Candidate 27: fits the user's viewing habit
of emotional-novel interpretation and story entanglements.''
\end{promptbox}
\noindent
No novel of the invented name appears anywhere in the slate or the shown
history.  The interest claim itself is grounded: the shown window contains
two novel-recap captions and two dozen captions on emotional themes.
Gemini, by contrast, names all ten of its returned candidates with
accurate one-line descriptions.  On this pair the judge decided item
claims for Gemini and the trade-off argument for RLVR, the same split as
in Figure~\ref{fig:rationale-judge}.

\paragraph{Case 5: long-history improvement.}
User row 174437.  The $441$-item history shows a persistent interest in
chat-record screenshots and emotional-confession videos ($17$ history
captions are explicitly chat-record posts), and the observed positive is
another such video.  From the candidates alone ($k{=}0$), the RLVR model
misses it: the positive is not among the returned ten.  At $k{=}64$ and
$k{=}256$ the same model holds it at rank $2$.  The Base model drifts
instead of improving: rank $7$ at $k{=}0$, out of the returned ten at both
$k{=}64$ and $k{=}256$.  The Gemini reference ranks the positive first at
$k{=}64$ and demotes it to rank $7$ at $k{=}256$, where its top pick
switches to an outfit-change video of a celebrity who dominates one stretch
of the added history.  One user reproduces the three curves of
Figure~\ref{fig:scaling}b.

\paragraph{Case 6: an unresolved invalid output.}
User row 174401, 1.7B Base on the text view, anchor setting.  Stage 1
closes its reasoning block after $2{,}033$ tokens with a plausible interest
summary but never produces a ranking, and forced extraction then emits
\texttt{[2, 3, 8, 1, 2, 3, 8, 1, 2, 3]}: four indices looped to length
ten.  Four distinct indices remain after deduplication, the format
requires ten, and the example scores zero.  The observed positive,
candidate $3$, sits inside the loop, so this zero comes from a broken
interface rather than a missed target; the interface diagnostics are
reported separately so that the two can be told apart.

\FloatBarrier
\section{Artifacts and Compute}
\label{app:artifacts}

The \anonword code release linked in Section~\ref{sec:intro} contains the
benchmark and training-data builders, parsers, reward code, prompt templates,
model revisions, environment specifications, and commands for all reported
runs.  The benchmark, the post-training datasets, and the per-user result
files behind every reported number are released with it.  Every result file
records the rendering, split, $k$, $N$, display seed, decoder, and selected
checkpoint when applicable.  For models served through an external provider,
we record an immutable revision when the provider exposes one and otherwise
record the provider model alias used for the call.  The release excludes API
credentials, account identifiers, and machine paths, and the released data,
which derive from OpenOneRec-RecIF, remain subject to its terms.

Training and inference use bf16 on A100 GPUs; SFT, RFT, and RLVR train on
four GPUs, and OPD uses two, one for the student and one for the frozen
teacher.  Exact software versions are
recorded in each method's configuration and result metadata; OPD and RLVR use
verl 0.8.0.  Checkpoints are selected on the 400 development users, and
the selected checkpoint is then scored on the test users.  Each method's
supervision budget and retention counts are recorded in the released
configurations and data statistics.

\end{document}